\documentclass[12pt]{iopart}

\usepackage{iopams}

\expandafter\let\csname equation*\endcsname\relax

\expandafter\let\csname endequation*\endcsname\relax
\usepackage{comment}
\usepackage{amsmath}
\usepackage{cite} 
\usepackage{tikz}
\usepackage{graphicx}
\usepackage{soul}
\usetikzlibrary{matrix,shapes,arrows,positioning,chains} 
\usepackage{hyperref}
\usepackage{amsmath}
\usepackage{amsfonts}
\usepackage{amssymb,latexsym,mathrsfs}
\usepackage{graphicx} 
\usepackage{float}
\usepackage[capitalize]{cleveref} 
\usepackage{xcolor}

\newcommand{\moy}[1]{\langle #1 \rangle}

\newcommand\bZ {{\mathbb Z}}

\newcommand\ztwo {{\mathbb Z}_2}

\begin{document}
	
\title{Long-ranged spectral correlations in eigenstate phases}

\author{Mahaveer Prasad$^1$, Abhishodh Prakash$^{1,2}$, J. H. Pixley$^3$, Manas Kulkarni$^{1}$}

\address{$^1$ International Centre for Theoretical Sciences, Tata Institute of Fundamental Research, Bengaluru -- 560089, India}
\address{$^2$ Rudolf Peierls Centre for Theoretical Physics, University of Oxford, Oxford, OX1 3PU, UK}
\address{$^3$ Department of Physics and Astronomy, Center for Materials Theory, Rutgers University, Piscataway, NJ 08854 USA}
\ead{abhishodh.prakash@physics.ox.ac.uk (he/him/his)}
\date{\today}
	
\begin{abstract} 						
We study non-local measures of spectral correlations and their utility in characterizing and distinguishing between the distinct eigenstate phases of quantum chaotic and many-body localized systems. We focus on two related quantities, the spectral form factor and the density of all spectral gaps, and show that they furnish unique signatures that can be used to sharply identify the two phases. We demonstrate this by numerically studying three one-dimensional quantum spin chain models with (i) quenched disorder, (ii) periodic drive (Floquet), and (iii) quasiperiodic detuning. We also clarify in what ways the signatures are universal and in what ways they are not. More generally, this thorough analysis is expected to play a useful role in classifying phases of disorder systems. 
\end{abstract}
	
\maketitle

	
	
	\section{Introduction}
	From the behavior of fluids to the motion of electrons in a crystal to the growth patterns of bacterial colonies, understanding the complex behavior of an extremely large number of interacting entities is of great interest in several areas of physics. Condensed matter physics and statistical mechanics have been extremely successful in explaining why the same underlying degrees of freedom can organize themselves in drastically different ways when their interactions are changed, despite the fact that quantum mechanical rules governing the interactions between them are extremely complicated and in most cases intractable.
	
	Although the study of equilibrium phases has been very successful~\cite{Plischke1994equilibrium,LandauLifshitz2013statistical}, understanding how thermal equilibrium itself emerges as an effective description of isolated physical systems is relatively underdeveloped although enormous progress \cite{Srednicki_ETH_PhysRevE.50.888,Deutsch_ETH_1991_PhysRevA.43.2046,BerryTabor1977_level,Haake_QuantumChaosBook} has been made, especially in recent years\cite{MaldacenaShenkerStanford2016bound,CurtVonKeyserlingk_Operatorspreading_PhysRevX.8.021013,NahumOPeratorSpreading_PhysRevX.8.021014,BertiniProsen_PhysRevLett.121.264101}. An important milestone in this direction is the formulation of the eigenstate thermalization hypothesis (ETH)~\cite{Srednicki_ETH_PhysRevE.50.888,Deutsch_ETH_1991_PhysRevA.43.2046}. ETH postulates the conditions under which an isolated quantum system equilibrates and can be described by quantum statistical mechanics. Systems that satisfy ETH are said to be quantum chaotic or ergodic whose eigenstates are highly i.e. ``volume-law'' entangled~\cite{kim2014testing} and eigenvalues are correlated due to level repulsions. They transport charge and energy rapidly, and their dynamics is ergodic in the sense that initial conditions are washed out at long times. In recent years, as suggested by increasing numerical and experimental evidence in conjunction with theoretical and phenomenological investigations, quantum many-body systems in the presence of strong quenched disorder (or quasiperiodicity) have emerged as a robust setting where ETH is violated. Such systems are said to be many-body-localized (MBL)~\cite{HuseNandkishore2015_MBLreview,abanin2019colloquium,gopalakrishnan2020dynamics,deng2017many,Imbrie_ProofMBL}  and are believed to form an ``eigenstate" phases of matter~\cite{HuseEtAl_LocalizationProtectedQuantumOrder_PhysRevB.88.014206,PotterVasseurParameswaran_eigenstatephases,PotterVasseurParameswaran_criticalglass_PhysRevLett.114.217201,AP_S3MBL_PhysRevB.96.165136} that are fundamentally distinct from ergodic systems.  In contrast to ergodic systems, the MBL phase is characterized by eigenstates with short-range  ``area law'' entanglement,  an absence of level repulsion in the energy eigenvalues, no charge or energy transport, and a retention of the memory of the initial conditions even at long times. Recent experiments on ultra-cold atomic gases~\cite{schreiber2015observation,choi2016exploring,lukin2019probing}, trapped ions~\cite{Smith_2016_MBLExpt}, superconducting qubits~\cite{roushan2017spectroscopic,xu2018emulating} and nuclear spins~\cite{wei2018exploring} have provided evidence for the existence of the MBL phase.

	It is therefore useful to have probes that can distinguish between various types of eigenstate phases. While conventional quantum phases of matter in equilibrium
are characterized by the properties of their ground states and low lying excitations, eigenstate order, in contrast, is characterized by the structure of the many-body spectrum. For ergodic systems, remarkably, eigenvalue correlations are reproduced by an appropriate random matrix ensemble~\cite{Haake_QuantumChaosBook}. These are manifested in local probes such as the Wigner-Dyson distribution of spectral gaps as well as in nonlocal probes such as the spectral form factor (SFF) which shows a robust, universal linear `ramp', which has been the subject of many recent studies ~\cite{Haake_QuantumChaosBook,Cotler_SFFChaos2017,ShenkerGharibyan2018onsetofRM,Liu_SFFChaos_PhysRevD.98.086026,ChenLudwig_PhysRevB.98.064309,BertiniProsen_PhysRevLett.121.264101,KosLjubotinaProsen_PhysRevX.8.021062}. MBL systems, on the other hand, exhibit an emergent integrability characterized by an extensive number of local integrals of motion and an energy spectrum that behaves like numbers drawn from a Poisson process~\cite{SerbynMoore_PhysRevB.93.041424,OganesyanHuse_2007_PhysRevB.75.155111,LevelstatistticsMBL_PhysRevB.99.104205,RMT_MBL_PhysRevLett.122.180601,RMT_MBL_PhysRevB.101.104201}. While Poisson signatures in MBL systems have been studied in local probes such as the distribution of nearest-neighbor spectral gaps and the value of the adjacent gap ratio~\cite{OganesyanHuse_2007_PhysRevB.75.155111}, recently, it was also shown that long-range spectral probes that measure correlations across the spectrum, such as the SFF, could also exhibit a unique
scaling form in the MBL phase~\cite{PrakashPixleyKulkarni_MBL_SFF_2021_PhysRevResearch.3.L012019,riser2020nonperturbative}. Interestingly, this has also been shown to be present in integrable quantum-mechanical systems such as integrable billiards~\cite{riser2020power}. 
	
	In this work, we investigate the utility of long-range spectral probes in characterizing and distinguishing between the MBL and ergodic phases further. First, we extend the results of our previous work in Ref.~\cite{PrakashPixleyKulkarni_MBL_SFF_2021_PhysRevResearch.3.L012019} where an analytical form for SFF was obtained and verified for energy conserving systems with quenched disorder. Here, we show that the SFF form is also applicable to quasiperiodic systems (deterministic potentials) and Floquet systems (which do not conserve energy). Second, we study a quantity related to the Fourier transform of the SFF, the distribution of all spectral gaps that we refer to as the density of gaps (DOG), for which we obtain an analytical form for Poisson numbers and random matrices (both of which have not been presented in the literature previously to the best of our knowledge).
	Using this, we show that the DOG also exhibits distinct signatures in the MBL and ergodic phases and also furnishes a means of tracking the transition between the two phases. Finally, we also clarify the robustness and universality of these spectral signatures in the MBL phase.  We show that unlike the `ramp' which is seen in the SFF of ergodic systems, the scaling form of the SFF in MBL systems depends on global aspects of the spectrum as a consequence of the lack of intrinsic correlations in MBL spectra. Altogether, our results show that the universal signatures present in long-range spectral probes that arise from both intrinsic correlations and global features of the spectrum can comprehensively characterize eigenstate phases more efficiently on finite-size simulations than either intrinsic or global only.
	
	The remainder of the paper is organized as follows. In \cref{sec:Measures of spectral correlations}, we review various measures of spectral correlations used in studying eigenstate phases and also provide analytical expressions for the form of the SFF and DOG in the MBL and ergodic phases calculated using Poisson numbers and random matrix eigenvalues, respectively. These analytical forms are numerically verified in \cref{sec:Numerical} using various one-dimensional models hosting MBL and ergodic phases. In \cref{sec:Origin}, we discuss the universality and robustness of these spectral signatures to various deformations of the spectrum and probes. Finally, in \cref{sec:Other}, we discuss the connections with other recent works before presenting our concluding remarks. Various additional details not present in the main text are relegated to the appendices.
	
	\section{Measures of spectral correlations}
	\label{sec:Measures of spectral correlations}
	In this section, we introduce the spectral probes that are going to be the main focus of our study. We discuss its expected behavior in ergodic and MBL systems by modeling their spectra using random matrix eigenvalues and Poisson numbers, respectively. In the following sections, we test these expectations numerically on various physical models. To orient the reader, we first review a well-known local probe - the adjacent gap ratio [see \cref{eq:r}] and then proceed to discuss the spectral form factor [see \cref{eq:SFF}] and density of all gaps [see \cref{eq:DOG}]. 
	
	\subsection{The adjacent gap ratio and local probes}
	Spectral correlations have been very useful in characterizing the ergodic properties of quantum systems. One of the first measures was by Wigner, who postulated that the repulsion of nuclear energy levels can be observed in the distribution of nearest-neighbor spectral gaps~\cite{mehta2004random} whose form he famously surmised. The Wigner surmise is also found to be applicable to all ergodic/quantum chaotic systems that are characterized by level repulsions. In contrast, in the absence of level repulsions such as in integrable systems, Berry and Tabor conjectured~\cite{BerryTabor1977_level} that the spectrum resembles a Poisson process and that the distribution of nearest-neighbor spectral gaps is exponential. This has also served as a good characterization of MBL systems whose spectrum resembles that of an integrable system~\cite{SerbynMoore_PhysRevB.93.041424,Imbrie2017LocalIO}. From this it is clear that the presence or absence of level repulsions can distinguish ergodic systems from MBL, and this can be probed locally in the spectrum. 
	
	A related local probe is the adjacent gap ratio \cite{OganesyanHuse_2007_PhysRevB.75.155111} defined in terms of successive nearest-neighbor gaps, $\delta_i,~\delta_{i+1}$ as follows.
	\begin{equation}\label{eq:r}
		r_i = \frac{\mathrm{min}(\delta_i,\delta_{i+1})}{\mathrm{max}(\delta_i,\delta_{i+1})}.
	\end{equation}
	It has been shown that the statistics of $r_i$ can characterize and distinguish between MBL and ergodic systems~\cite{OganesyanHuse_2007_PhysRevB.75.155111}. For example, tracking the mean value $\moy{r}$ can tell us if the system is chaotic or integrable/many-body localized and estimate the location of the transition between them. If the system is in the MBL phase, irrespective of symmetries present, $\langle r \rangle =2\log 2-1 \approx 0.39$. For ergodic systems, the value of $\moy{r}$ depends on the global symmetries present in the system. For example, when time-reversal (and spin rotational) symmetry is present $\moy{r} \approx 0.53$ and when no symmetries are present, $\moy{r} \approx 0.6$. We remark that the statistics of the adjacent gap ratio can be estimated very well from the distribution of nearest-neighbor gaps using random matrix ensembles and Poisson numbers (see Refs.~\cite{Haake_QuantumChaosBook,mehta2004random,OganesyanHuse_2007_PhysRevB.75.155111,AtasBogomolnyGiraudRoux_adjacentgapratioFormula_PhysRevLett.110.084101} for details, also reviewed in \ref{app:r}). At this stage it is natural to ask whether nonlocal probes of spectral correlations, which probe larger energy scales beyond nearest-neighbor gaps, also carry unique signatures of quantum chaos and MBL. This is indeed true and has been the subject of several past and recent works~\cite{Haake_QuantumChaosBook,PrakashPixleyKulkarni_MBL_SFF_2021_PhysRevResearch.3.L012019,Cotler_SFFChaos2017,ShenkerGharibyan2018onsetofRM,Liu_SFFChaos_PhysRevD.98.086026,ChenLudwig_PhysRevB.98.064309,BertiniProsen_PhysRevLett.121.264101,KosLjubotinaProsen_PhysRevX.8.021062}, including this paper.

	\subsection{The spectral form factor (SFF)}
	\label{sec:SFF_Chaos}
	An important non-local spectral probe is the spectral form factor (SFF), which we denote as $K(\tau,N)$.  SFF and its connected version (CSFF) $K_c(\tau,N)$ are defined in terms of ensembles of $N$ eigenvalues $\{E_i\}$ as follows~\cite{Haake_QuantumChaosBook}
	\begin{align}
		K(\tau,N) &=  \langle \sum_{m,n=1}^N e^{i \tau (E_m - E_n)} \rangle, 
		\label{eq:SFF}\\
		K_c(\tau,N) &=  \langle \sum_{m,n=1}^N e^{i \tau (E_m - E_n)}   \rangle - |\langle \sum_{m=1}^N e^{i \tau E_m}   \rangle|^2,
		\label{eq:SFFC}
	\end{align}
	where $\langle \dots \rangle$ stands for average over ensembles. 
	To reduce clutter, we will focus on the SFF [\cref{eq:SFF}] but comment on the CSFF [\cref{eq:SFFC}] when necessary. For a fixed number $N$ (which we assume to be large throughout this paper), as $\tau$ is tuned, the SFF probes the correlations in the spectrum on scales inversely proportional to $\tau$. It is useful to separate the behavior of the SFF on various $\tau$ scales. In general, there are three $\tau$ regions separated by the so-called Thouless time $\tau_{T}\sim \frac{1}{\mu N}$ and Heisenberg time $\tau_H \sim \frac{1}{\mu}$ where $\mu$ is the mean level spacing \cite{Haake_QuantumChaosBook,PrakashPixleyKulkarni_MBL_SFF_2021_PhysRevResearch.3.L012019}: 
	\smallskip 
	\begin{enumerate}
		\itemsep1em 
		\item \textit{Early $\tau$}: For small values of $\tau << \tau_{T}$, the SFF probes the spectrum on the bandwidth scale and is sensitive to the tails of the spectrum.
		
		\item \textit{Intermediate $\tau$}: For $\tau_{T}<\tau<\tau_H$, the behavior is expected to be dominated by universal correlations, if present. This is usually the regime of prime interest.
		
		\item \textit{Late $\tau$}: For large values of $\tau >> \tau_H$, the SFF probes the spectrum on the scale of the mean level spacing where the levels are quantized. In the absence of accidental degeneracies, the expression of SFF~\cref{eq:SFF} in this regime is thus dominated by terms where $E_m = E_n$ and the SFF plateaus at $K \approx N$.
		
	\end{enumerate}

It is interesting to note that SFF also has a broader appeal. In addition to encoding information about long range correlations, SFF is also a highly valuable computational quantity. Some of the reasons are, (i) SFF is closely connected to a dynamical quantity called the survival probability~\cite{Herrada2023,daug2023many}. (ii) SFF is amenable to analytical calculations for certain systems where other quantities are far from being analytically tractable. For e.g. there has recently been work on hydrodynamic theory of the connected spectral form factor~\cite{Winer2022}. (iii) There are deep analytical insights for SFF results for eigenvalues of random matrix theories~\cite{Haake_QuantumChaosBook,Cotler_SFFChaos2017,ShenkerGharibyan2018onsetofRM,Liu_SFFChaos_PhysRevD.98.086026,ChenLudwig_PhysRevB.98.064309,BertiniProsen_PhysRevLett.121.264101,KosLjubotinaProsen_PhysRevX.8.021062}. This makes it possible to explore deep connections between chaotic quantum Hamiltonians and random matrix theory through the lens of SFF. 
	
	\subsubsection{The spectral form factor for random matrices}
	\begin{figure}[!h]
		\centering
		\begin{tabular}{c}
			\includegraphics[width=0.45\textwidth]{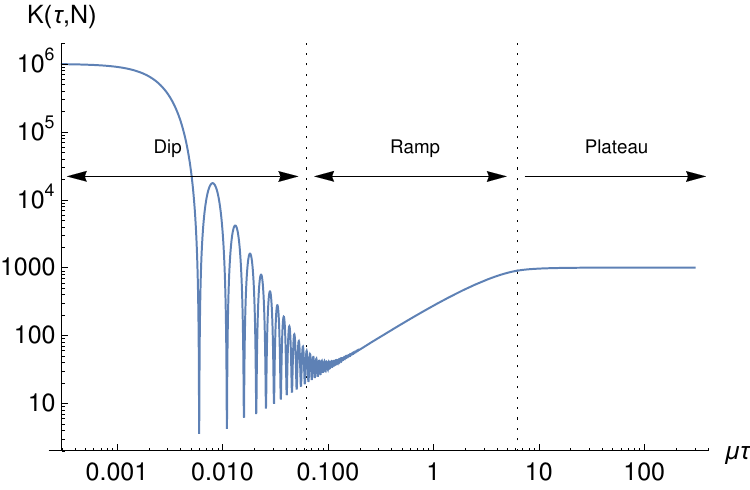} 
		\end{tabular}
		\caption{SFF for GOE ensemble  shown in ~\cref{eq:SFF_Chaos} \label{fig:SFF_Chaos} with N=1000. The separation between dip, ramp, and plateau regimes is schematically indicated using dotted lines.}
	\end{figure}
	
	The SFF for random matrix spectra, applicable to quantum chaotic systems, has been the subject of intense study and has recently attracted renewed interest~\cite{Haake_QuantumChaosBook,PrakashPixleyKulkarni_MBL_SFF_2021_PhysRevResearch.3.L012019,Cotler_SFFChaos2017,ShenkerGharibyan2018onsetofRM,Liu_SFFChaos_PhysRevD.98.086026,ChenLudwig_PhysRevB.98.064309,BertiniProsen_PhysRevLett.121.264101,KosLjubotinaProsen_PhysRevX.8.021062}. We briefly review the results here. The precise nature of the random matrix SFF depends on the underlying symmetries~\cite{Haake_QuantumChaosBook,mehta2004random}. For systems with time-reversal symmetry, the appropriate RMT ensemble is the Gaussian Orthogonal Ensemble (GOE) for which the approximate expression of SFF can be written as~\cite{Haake_QuantumChaosBook,Cotler_SFFChaos2017,ShenkerGharibyan2018onsetofRM,Liu_SFFChaos_PhysRevD.98.086026}
	\begin{align}
		\label{eq:SFF_Chaos}
		K^{GOE}(\tau,N) &=  K^{GOE}_c(\tau,N)+  \left[\frac{\pi}{\mu \tau} J_1 \left(\frac{2 N \mu \tau}{\pi}\right)\right]^2 , \\
		K^{GOE}_c(\tau,N ) &= N\begin{cases}
			\frac{\mu \tau}{\pi} -\frac{\mu \tau}{2\pi} \log \left(1+ \frac{\mu \tau}{\pi}\right) ~~ 0 <\mu \tau < 2 \pi \\
			2-  \frac{\mu \tau}{2 \pi} \log \left( \frac{\mu \tau + \pi}{\mu \tau -\pi}\right) ~~~2 \pi<\mu  \tau < \infty 
		\end{cases} ,\nonumber
	\end{align}
	where $J_1(x)$ is the Bessel function of the first kind. A plot of \cref{eq:SFF_Chaos} is shown in \cref{fig:SFF_Chaos} where we can see three qualitative regimes - an early dip ($\tau < \tau_{T}$), intermediate ramp ($\tau_{T}<\tau<\tau_H$) and a late saturation ($\tau > \tau_H$).  The SFF for other RMT ensembles also exhibit these three regimes, which are considered to be universal features of level repulsions and many-body quantum chaos.

	\subsubsection{The spectral form factor for Poisson numbers}
	\label{sec:SFF_Poisson}
	\begin{figure}[!h]
		\centering
		\begin{tabular}{cc}
			\includegraphics[width=0.45\textwidth]{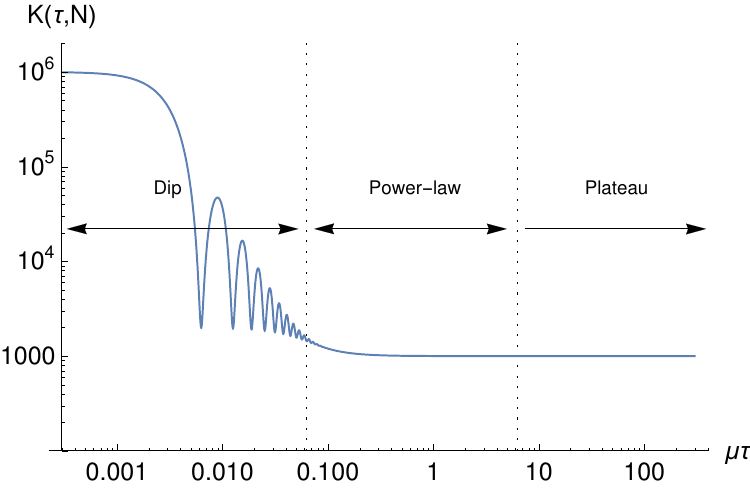} &
			\includegraphics[width=0.45\textwidth]{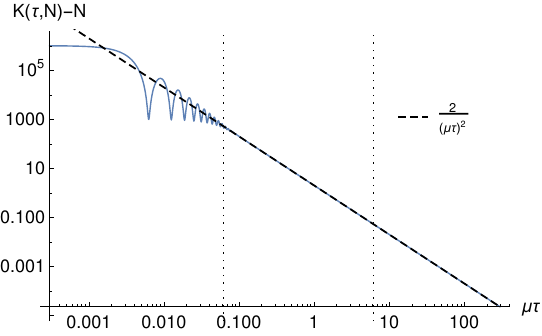} 
		\end{tabular}
		\caption{{Left: SFF for Poisson levels shown in \cref{eq:SFF_Poisson}. Right: The power-law scaling form for $\tau_T < \tau < \tau_H$ is exposed by subtracting the saturation value of N (chosen to be 1000 for both plots). \label{fig:SFF_Poisson}}}
	\end{figure}
	
	The SFF for Poisson numbers applicable to MBL and integrable systems was investigated in detail only recently \cite{PrakashPixleyKulkarni_MBL_SFF_2021_PhysRevResearch.3.L012019,riser2020nonperturbative,riser2020power}. An uncorrelated spectrum that resembles that of integrable and MBL systems can be generated~\cite{PrakashPixleyKulkarni_MBL_SFF_2021_PhysRevResearch.3.L012019} by starting with nearest-neighbour gaps $\{\delta_n\}$ from an exponential distribution. 
	\begin{equation}
		P(\delta_n) = \frac{1}{\mu} e^{-\delta_n/\mu},
	\end{equation}
	and summing them up
	\begin{equation}
		E_n = \sum_{r=1}^n \delta_n. \label{eq:built_spectrum}
	\end{equation}
	The joint 2-point probability distribution for this spectrum,  $P(E_n,n;E_m,m)$ i.e. the probability that the $m^{th}$ eigenvalue is $E_m$ and the $n^{th}$ eigenvalue is $E_n$ is 
	\begin{align}
		P(E_n,n;E_m,m) = p(E_n,n)~ p(E_m-E_n,m-n), \label{eq:Poisson_2pt}
	\end{align}
	where $p(E_k,k)$ ($k =1,2,3,\ldots$) is the well-known Poisson distribution.
	\begin{equation}
		\label{eq:Poisson distribution}
		p(E_k,k) =  \begin{cases}
			\frac{e^{-\frac{E_k}{\mu} }}{\mu (k-1)!} \left(\frac{E_k}{\mu}\right)^{k-1}~E_k \ge 0 \\
			~~~~~~~~0 ~~~~~~~~~~~~~E_k <0
		\end{cases} 
	\end{equation}
	Using this, the SFF can be calculated to obtain the following expression (see Ref.~\cite{riser2020nonperturbative} and the supplementary materials of Ref.~\cite{PrakashPixleyKulkarni_MBL_SFF_2021_PhysRevResearch.3.L012019} for the derivation).
	\begin{align}
		K^P(\tau,N) &= N + \frac{2}{(\mu\tau)^2}  - \frac{ (1+i \mu \tau)^{1-N} + (1-i \mu \tau)^{1-N}  }{(\mu\tau)^2}. \label{eq:SFF_Poisson} 
	\end{align}
	In \cref{fig:SFF_Poisson}, we see that the SFF for the spectrum in \cref{eq:built_spectrum} can also be divided into three $\tau$ regimes. Now we focus on the intermediate $\tau$ regime $\frac{1}{N}< \mu \tau < 1$ when the SFF form reduces to~\cite{PrakashPixleyKulkarni_MBL_SFF_2021_PhysRevResearch.3.L012019}
	\begin{align}
		K^P(\tau,N) &= N + \frac{2}{(\mu \tau)^2} + \ldots \label{eq:KPoisson_universal}
	\end{align}
	We can see that if we subtract the asymptotic value of $N$, $K(\tau,N) - N$ (which we will refer to as the reduced SFF) assumes a power law form independent of $N$ with a fixed exponent and can be used to characterize systems with uncorrelated spectra as shown in Ref.[\cite{PrakashPixleyKulkarni_MBL_SFF_2021_PhysRevResearch.3.L012019}]. Although the spectrum in \cref{eq:built_spectrum} was built in a specific way, it captures many essential features of uncorrelated levels. For example, if $N_R$ numbers were drawn from any distribution and $N$ of those were selected from a fixed window after ordering, the distribution of $k^{th}$ neighbour spectral gaps $E_n - E_{n+k}$ of these $N$ numbers approaches the Poisson distribution \cref{eq:Poisson distribution} (see the supplementary materials of Ref.\cite{PrakashPixleyKulkarni_MBL_SFF_2021_PhysRevResearch.3.L012019}). Similarly, the same is expected to be true for $N$ energy levels chosen from the middle of the spectrum of a system deep in the MBL phase. However, one feature of the spectrum in \cref{eq:built_spectrum} is the fact that all levels $E_n$ are positive definite, which results in a sharp step-like feature in the density of states (DOS) which is absent if we choose $N$ levels from the middle of a quantum many-body spectrum. As we will see in the coming sections, signatures such as the one shown in \cref{eq:KPoisson_universal} are unaffected by this DOS feature and describe the spectra of MBL systems fairly accurately. Throughout the paper, we will focus only on the SFF \cref{eq:SFF}. Some subtleties regarding connected SFF, especially regarding the effects of sharp DOS \cref{eq:SFFC} will be discussed in \ref{app:ConnectedSFF}.
	
	\subsection{The density of all gaps (DOG)}
	We now discuss another useful quantity --- the density of all gaps (DOG), $\chi(x,N)$, defined as 
	\begin{equation} 
		\chi(x,N) = \frac{1}{N (N-1)} \big\langle \sum_{m \neq n=1}^N \delta(x-(E_m - E_n))  \big\rangle \label{eq:DOG}.
	\end{equation}
	and is related to the SFF in ~\cref{eq:SFF} as follows
	\begin{align}
		K(\tau,N) &=  N + N(N-1) \int_{-\infty}^\infty dx~ e^{i \tau x}~ \chi(x,N).
	\end{align}
	In other words, the SFF is related to the Fourier transform of the DOG. We note that often the SFF is written in terms of the two-point density correlator~\cite{mehta2004random}
	\begin{align}
		K(\tau,N) &= N +  N(N-1)\\& \int_{-\infty}^\infty dx \int_{-\infty}^\infty dy~ e^{i \tau (x-y)}~ \rho^{(2)}(x,y,N).\\
		\rho^{(2)}(x,y,N) &= \frac{1}{N (N-1)} \big\langle \sum_{m \neq n=1}^N \delta(x-E_m) \delta(y-E_n) \big\rangle.
	\end{align}
	We can relate $\chi(x,N)$ in \cref{eq:DOG} to $\rho^{(2)}(x,y,N)$ as follows
	\begin{equation}
		\chi(x,N) = \int_{-\infty}^\infty du ~\rho^{(2)}(x+u,u,N). \label{eq:chi rho2}
	\end{equation}
	We will see that studying the DOG exposes interesting details about spectral correlations. 
	
	\begin{figure}[h]
		\centering
		\includegraphics[width=0.49\textwidth]{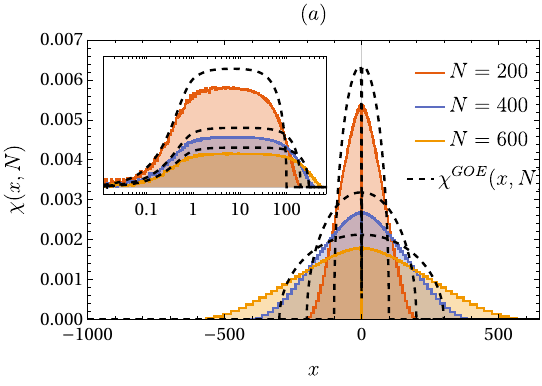} \hfill
		\includegraphics[width=0.49\textwidth]{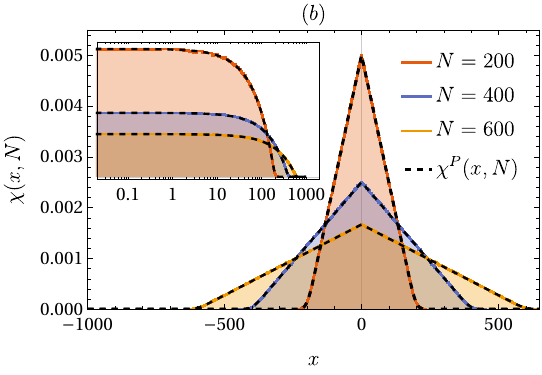}
		\caption{{(a) DOG [\cref{eq:DOG}] for the eigenvalues of GOE  matrices  compared with analytical expression in  \cref{eq:chi_RMT}. (b) DOG [\cref{eq:DOG}] of Poisson levels compared with analytical expression in \cref{eq:chi_poisson1}. The insets show the same figure restricted to positive values with the x-axis in log scale to clarify the behavior near the origin. The plot for each $N$ is generated using $5000$ levels with mean level spacing $\mu=1$. \label{fig:DOG_poisson_RMT}}}
	\end{figure} 
	
	\subsubsection{Density of all gaps for random matrices}
	
    We can obtain approximate analytical forms for the DOG for both the RMT and Poisson cases. Let us start with the former. The expression for GOE matrices is ~\cite{mehta2004random,Liu_SFFChaos_PhysRevD.98.086026,Cotler_SFFChaos2017,ChenLudwig_PhysRevB.98.064309}
    \begin{equation}
        \chi^{GOE}(x,N) = \rho(x,N) (1- Y_2(x)) \label{eq:chi_RMT}
    \end{equation}
where $\rho(\lambda,N)$ is the well-known semi-circle density of states
\begin{equation}
	\rho(\lambda,N) = \begin{cases}
		\frac{4}{\pi N \mu }  \sqrt{1 - \left(\frac{2 \lambda}{\mu N}\right)^2} &|\lambda| <\frac{N \mu}{2}\\
		~~~~~~~~~~~~0~~~~~~~~~~~~ &|\lambda|>\frac{N \mu}{2}
	\end{cases}.
\end{equation}
and $Y_2(x)$  is the asymptotic connected two-point spectral correlator~\cite{mehta2004random}
\begin{equation}
	Y_2(\mu r) = \left(\frac{\sin(\pi r)}{\pi r}\right)^2 \\+ \left[\int_r^\infty ds  \left(\frac{\sin(\pi s)}{\pi s}\right)\right] \left[\frac{\partial }{\partial r}  \left(\frac{\sin(\pi r)}{\pi r}\right)\right]. 
\end{equation}
Following \cite{ChenLudwig_PhysRevB.98.064309,Liu_SFFChaos_PhysRevD.98.086026,Cotler_SFFChaos2017}, we improve $Y_2(x)$ by introducing a density correction to the argument. \Cref{fig:DOG_poisson_RMT}(a) shows a comparison between the numerically computed DOG and the expression in \cref{eq:chi_RMT}. The agreement is very good for small gaps $x$ that corresponds to $\tau > \tau_T$ in the SFF but deviates for large gaps $x$ that correspond to $\tau < \tau_T$ in the SFF.

	\subsubsection{Density of all gaps for Poisson numbers}
For Poisson spectra generated as described in ~\cref{sec:SFF_Poisson}, we can compute $\chi(x,N)$ exactly using the distribution in \cref{eq:Poisson distribution} in the formula \cref{eq:DOG} to get 
\begin{equation}
	\chi^{P}(x,N)
	= \frac{\left( e^{- \frac{|x|}{\mu}} \left( \frac{|x|}{\mu}\right)^N - \left(\frac{|x|}{\mu} - (N-1) \right) \Gamma\left(N,\frac{|x|}{\mu} \right)  \right)}{\mu (N-1) N!}, 
	\label{eq:chi_poisson1}
\end{equation}
where $\Gamma(x,N)$ is the incomplete Gamma function defined as follows (for integer N)
\begin{equation}
	\Gamma(N,x) = \int_x^\infty dt ~t^{N-1} e^{-t} = e^{-x} (N-1)! \sum_{k=0}^N \frac{x^k}{k!}.
\end{equation}
As seen in \cref{fig:DOG_poisson_RMT}, the exact expression perfectly matches the numerical data.

	\label{sec:chi0_orderparameter}
	Much like the adjacent gap ratio, the DOG is a useful diagnostic not only for characterizing the MBL and ergodic phases, but also for estimating the location of the transition between them. For example, from the expressions \cref{eq:chi_RMT,eq:chi_poisson1}, we see that for RMT levels, $\chi(0,N) \rightarrow 0$ while for Poisson numbers $\chi(0,N) \rightarrow \frac{1}{N}$. Thus, the quantity $N \chi(0,N)$ takes values between 0 and 1 and tracking it on finite-size systems should give us an estimate of the transition between the MBL and ergodic phases where different system sizes cross. We will see in the next section that this is indeed true for various physical models hosting an MBL to ergodic transition.

	\section{Numerical study of physical models}
	\label{sec:Numerical}
\begin{figure*}[!tbp]
		\centering
		\includegraphics[width=0.32\textwidth]{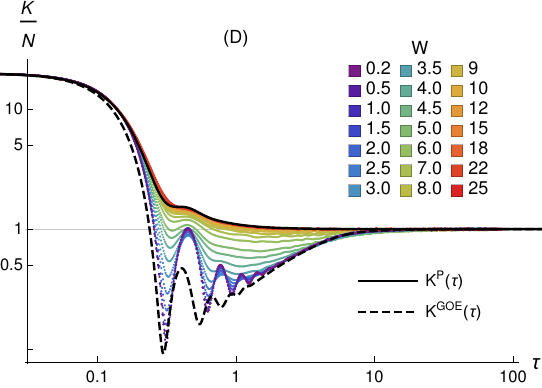} 
		\includegraphics[width=0.32\textwidth]{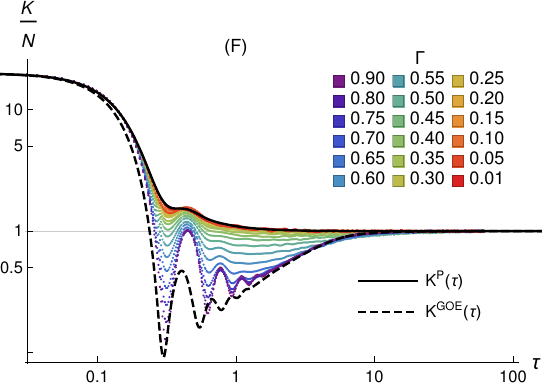} 
		\includegraphics[width=0.32\textwidth]{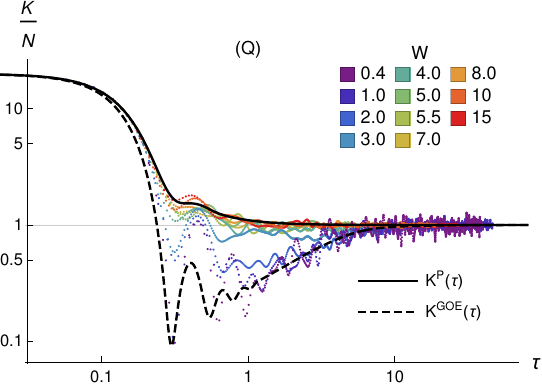}
		\includegraphics[width=0.32\textwidth]{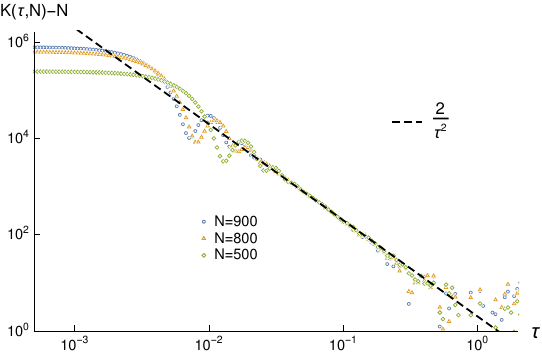}
		\includegraphics[width=0.32\textwidth]{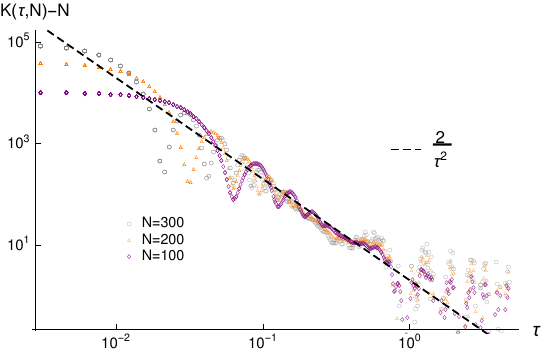}
		\includegraphics[width=0.32\textwidth]{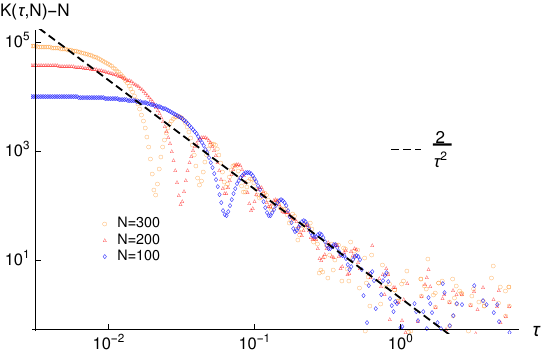}
		\caption {{ Top row: SFF [\cref{eq:SFF}] for $N=20$ levels across  various disorder strengths. Bottom row: Reduced SFF deep in the MBL phase with power-law scaling. Plots in left, middle and right columns correspond to the disordered spin chain (D), Floquet spin chain (F) and quasi-periodic spin chain (Q) defined in \cref{eq:Hamiltonian,eq:Floquet unitary,eq:H_QP} respectively. System sizes used are (from left to right) $L=14, 10,12$ for the top row and $L=18,12,16$ for the bottom row. The disorder strengths used for the bottom row is (from left to right) $W=25,~\Gamma = 0.01$ and $W =15$.}}
		\label{fig:SFF_numerics}
	\end{figure*}

	We now consider various quantum-mechanical many-body systems that are known to host an MBL to an ergodic transition and numerically study their spectrum focusing on the probes discussed in \cref{sec:Measures of spectral correlations} as follows:
	
	\emph{SFF}: We extract the spectral form factor $K(\tau,N)$ defined in \cref{eq:SFF} and compare it with the analytical form obtained from RMT shown in \cref{eq:SFF_Chaos} in the ergodic phase and to the form obtained from Poisson numbers shown in \cref{eq:SFF_Poisson} in the MBL phase. In particular, we verify that deep in the MBL phase we observe the power law scaling shown in \cref{eq:KPoisson_universal}. The SFF plots are shown in \cref{fig:SFF_numerics} for the various models. 
	
\begin{figure*}[!tbp]
		\centering
		\includegraphics[scale=0.355]{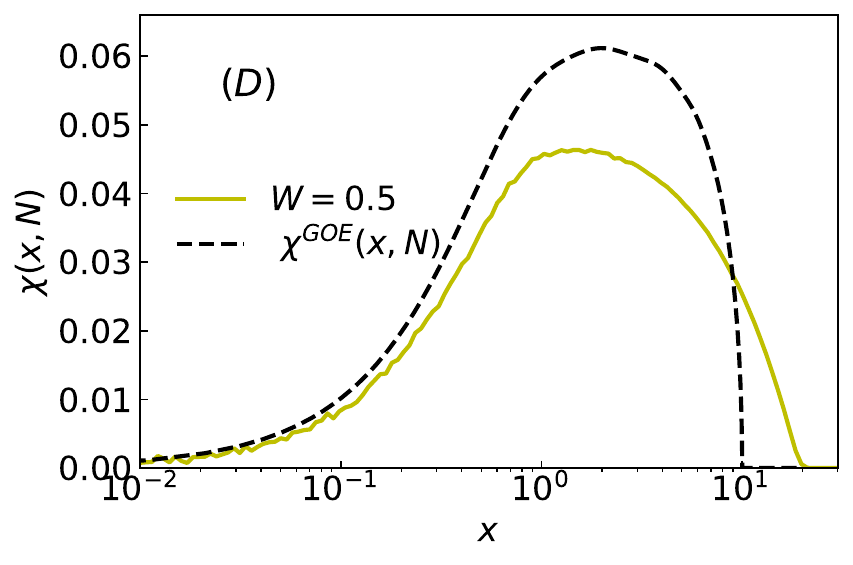}
		\includegraphics[scale=0.355]{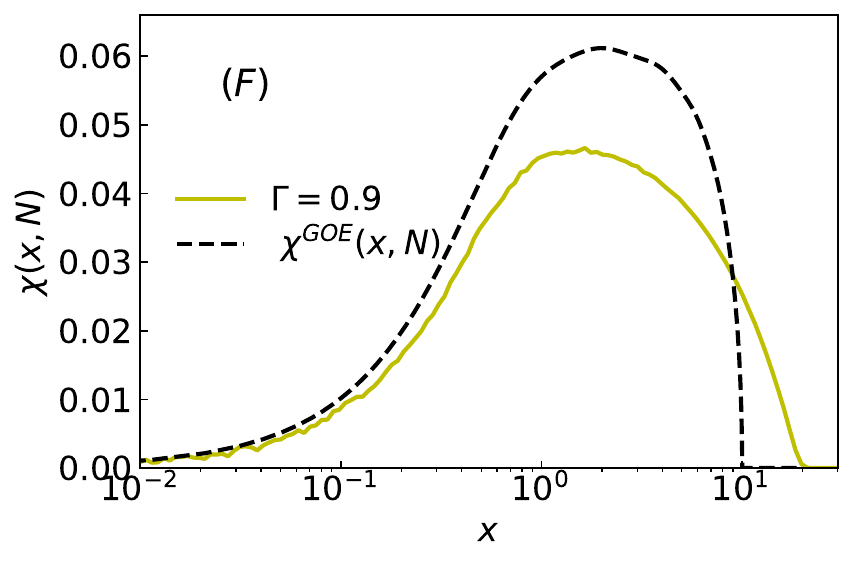}
		\includegraphics[scale=0.355]{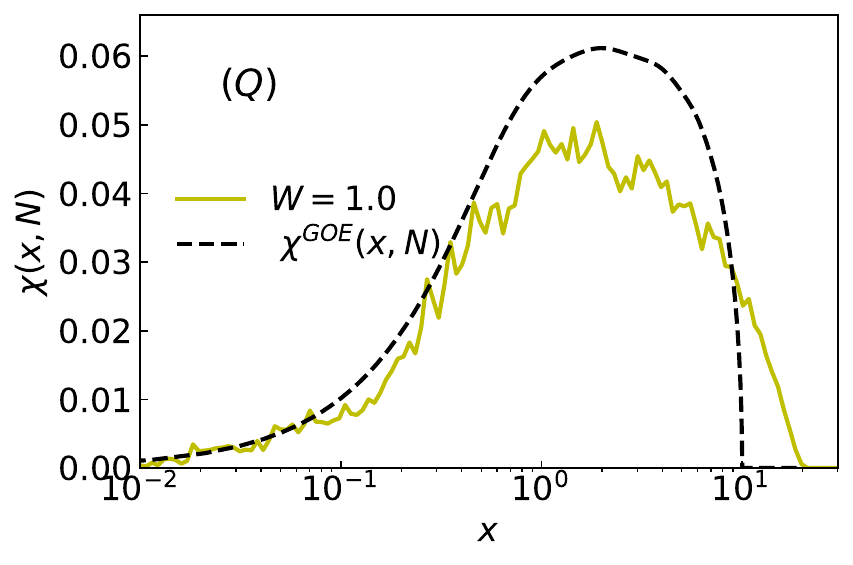}
		\includegraphics[scale=0.355]{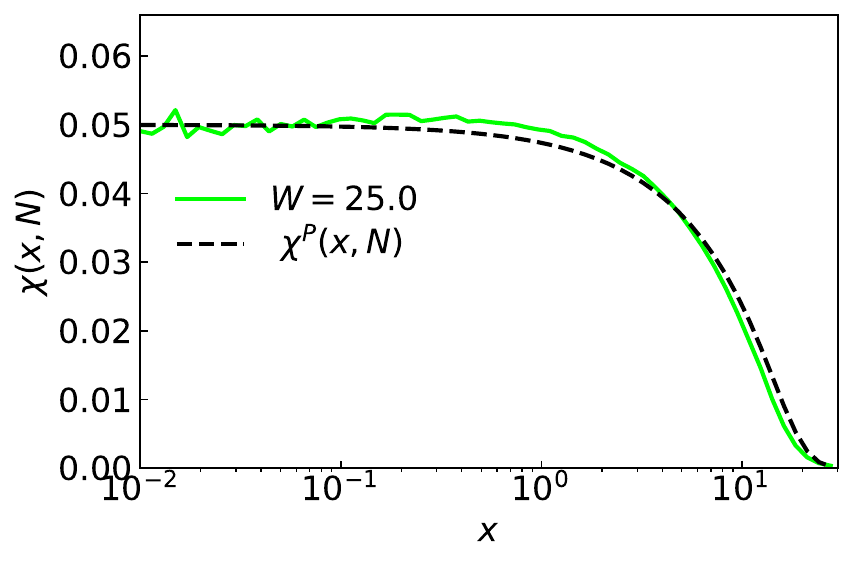}
		\includegraphics[scale=0.355]{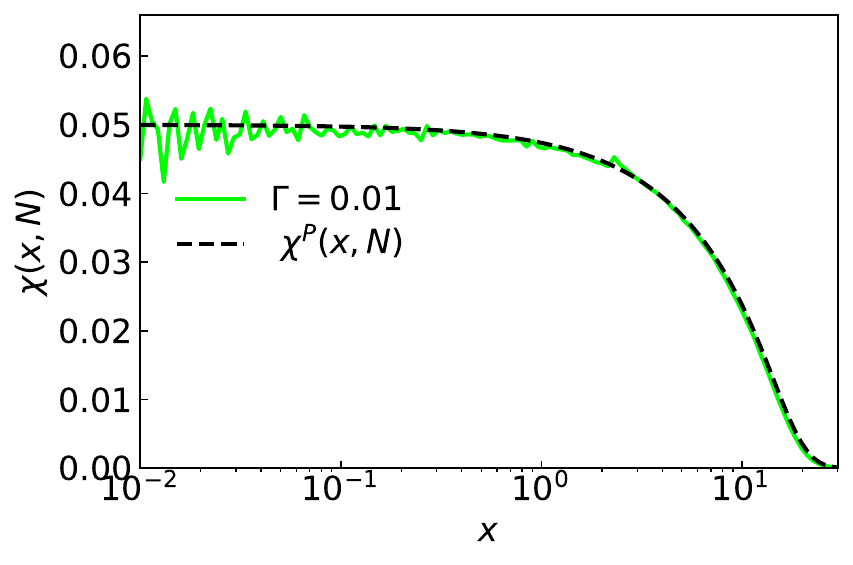}
		\includegraphics[scale=0.355]{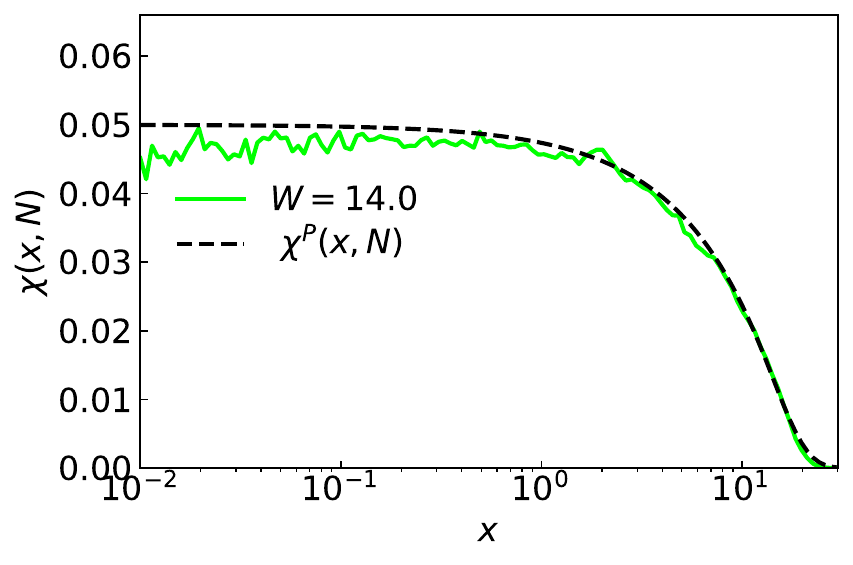}
		\caption {{  DOG [\cref{eq:DOG}] plots deep in the ergodic phase (top row) and the MBL phase (bottom row) for (from left to right) the disordered spin chain (D), the Floquet spin chain (F) and the quasiperiodic spin chain (Q) defined in \cref{eq:Hamiltonian,eq:Floquet unitary,eq:H_QP} respectively, compared to the analytical expressions in \cref{eq:chi_RMT,eq:chi_poisson1} (dotted lines). $N=20$ levels drawn from the middle of the spectrum and rescaled to set mean level spacing at unity, and plotted the numerical data with positive values of gaps with the log scale on x-axis.  The system sizes used are (from left to right) $L=12, 10,12$. }}
		\label{fig:DOG}
	\end{figure*}

	\emph{DOG}: We compute the density of all gaps defined in \cref{eq:DOG} and compare it with the analytical expression obtained from RMT shown in \cref{eq:chi_RMT} in the ergodic phase and to the form obtained from the Poisson spectra shown in \cref{eq:chi_poisson1} in the MBL phase. The DOG plots are shown in \cref{fig:DOG} for the various models. As discussed in \cref{sec:chi0_orderparameter}, we also use the appropriately normalized density of zero gaps $[N \chi(0,N)] \in [0,1]$ as an order parameter to track the transition between the MBL and the ergodic phases. This is shown in \cref{fig:r_chi0} where we compare the location of the transition determined by $[N \chi(0,N)]$ with the same determined by the more conventional adjacent gap ratio ($r$) defined in \cref{eq:r}. Discussions on the sensitivity of numerical computations of $\chi(x,N)$ to binning are discussed in \cref{sec:level_repulsion} and \ref{app:binning}
	
	In order to perform our numerical analysis, we employ the following prescription: From each disorder realization of the model, we extract $N$ consecutive levels from the middle of the spectrum where $N$ is smaller than the total number of levels $N_R$ and then proceed to compute the SFF and DOG. This does not produce any sharp DOS features of the kind present in \cref{eq:built_spectrum}. We also rescaled the data to set the mean level spacing $\mu$, averaged across disorder samples, to 1 for convenience. Whenever we have a global $U(1)$ symmetry, we extract the levels from the zero total magnetization sector, $\sum_j S_j^z=0$. Additional details of numerical analysis, such as the number of disorder samples the data is averaged over to produce all the figures, are tabulated in \ref{app:disorder} to reduce clutter. In general, all our numerical results are in excellent agreement with analytical predictions. We provide details of each model and comment on various features below. For clarity, a brief summary of the models considered and the location of their results are shown in Table~\ref{tab:table1}. 
	\\
	\begin{table}[!h]
		\begin{center}
			\begin{tabular}{|c|c|c|c|}
                \hline
				\textbf{Model} & \textbf{Details} & \textbf{SFF} &  \textbf{DOG }\\
				\hline
				Disordered  & \cref{eq:Hamiltonian} & \cref{fig:SFF_numerics} (left) & \cref{fig:DOG} (left)\\
				spin chain &  &  & \cref{fig:r_chi0} (bottom left) \\
				\hline
				Floquet  & \cref{eq:Floquet unitary} & \cref{fig:SFF_numerics} (middle)& \cref{fig:DOG} (middle)\\
				spin chain& &  & \cref{fig:r_chi0} (bottom middle)\\
				\hline
				Quasi-periodic & \cref{eq:H_QP} & \cref{fig:SFF_numerics} (right)& \cref{fig:DOG} (right) \\
				spin chain &  & \cref{fig:QP_oscillations} (right)  & \cref{fig:r_chi0} (bottom right) \\
				\hline
			\end{tabular}
		\end{center}
		\caption{Brief summary of the three models considered and the location of the main results of each. \label{tab:table1}}
	\end{table}
	
	\subsection{Spin chain Hamiltonian with quenched disorder}
	
	We begin with the quantum spin chain Hamiltonian with quenched disorder considered in Refs. \cite{PrakashPixleyKulkarni_MBL_SFF_2021_PhysRevResearch.3.L012019,vsuntajs2020quantum}.
	\begin{equation}
		\label{eq:Hamiltonian}
		H = \sum_i   J_1 (S^x_i S^x_{i+1} + S^y_i S^y_{i+1} + \Delta S^z_i S^z_{i+1})+  w_i S^z_i \\ +  \sum_i J_2 (S^x_i S^x_{i+2} + S^y_i S^y_{i+2} + \Delta S^z_i S^z_{i+2}). 
	\end{equation}
	Here and henceforth, $S^\alpha$ are spin-half operators that can be written in terms of Pauli matrices as $S^\alpha= \frac{1}{2} \sigma^\alpha$ and $w_i$ are random numbers drawn from a uniform distribution $w_i \in [-W, W]$. Following Refs.~\cite{vsuntajs2020quantum,PrakashPixleyKulkarni_MBL_SFF_2021_PhysRevResearch.3.L012019}, we set $J_1 = J_2 = 1.0$ and $\Delta = 0.55$. This Hamiltonian has $U(1)$ spin-rotation symmetry which allows us to work with the spectrum of the zero magnetization sector (half filling) for numerical analysis.
	Variants of this model have been previously studied~\cite{khemani2017critical,khemani2017two} and are known to have a thermal phase at weak disorder and an MBL phase in strong disorder, as can be seen by tracking the adjacent gap ratio ($r$) defined in \cref{eq:r}.
	As shown in \cref{fig:r_chi0} (top row, left column), tracking the adjacent gap ratio indicates that the strength of the critical disorder is somewhere near $W_c \approx 7.3$ where the curves cross for different sizes of the system. 
	\subsection{Floquet spin chain with quenched disorder}
	Local conservation laws produce diffusive hydrodynamic modes that can slow down dynamics and obscure the thermalization properties of the system \cite{kim2014testing}. Periodically driven Floquet systems with quenched disorder and no global symmetries serve as a useful testing ground for studying eigenstate phases because they contain no conservation laws, including energy. As a result, the system is allowed to thermalize rapidly in the ergodic phase, leading to a sharper MBL-ergodic transition. Now we consider the Floquet model defined in Ref.\cite{zhang2016floquet}. This is described by the following Floquet unitary operator that generates the time evolution for one time period, $2T$
	\begin{equation}
		U(T)=\exp\left(-i \frac{T}{2}H_x \right)\exp\left( {-iT} H_z \right)\exp\left(-i \frac{T}{2}H_x \right), \label{eq:Floquet unitary}
	\end{equation}
	where, the Hamiltonians $H_{x}$ and $H_{z}$ are defined as follows:
	\begin{equation}\label{eq:H_Fq}
		\begin{aligned}
			H_{x} &= g\Gamma\sum_{j=1}^{L} \sigma_{j}^{x},  \\
			H_{z} &=\sum_{j=1}^{L-1} \sigma_{j}^{z}\sigma_{j+1}^{z} +\sum_{j=1}^{L}(h+g \sqrt{1-\Gamma^2} G_{j}) \sigma_{j}^{z}.
		\end{aligned}
	\end{equation}
	
	Our choice of parameters $(g,h)=(0.9045, 0.8090)$  and the time period  $2T=1.6$ are the same as studied in Ref.~\cite{zhang2016floquet}. $\{G_{j}\}$ are independent Gaussian  random numbers with zero mean and unit standard deviation where it was argued that the critical point $\Gamma_{c}$ was near $0.3$. For $\Gamma< \Gamma_{c}$ this system is in MBL phase and for $\Gamma > \Gamma_{c}$ it is ergodic. Eigenvalues of unitary operator \cref{eq:Floquet unitary} are pure complex phases $\{e^{i \theta_j}\}$ where $\{\theta_j\}$ define the Floquet spectrum and take values on the unit circle. We will use $\{\theta_j\}$ to study the  SFF and DOG. 
	\begin{align}
		K(\tau,N) &=  \langle \sum_{m,n=1}^N e^{i \tau (\theta_m - \theta_n)} \rangle, \label{eq:SFF_theta}\\
		\chi(x,N) &= \frac{1}{N (N-1)} \big\langle \sum_{m \neq n=1}^N \delta(x-(\theta_m - \theta_n))  \big\rangle.
	\end{align} \label{eq:DOG_theta}
	Since $\{\theta_j\}$ are only well defined on a unit circle, for the SFF expression in \cref{eq:SFF_theta} to be well defined, $\tau$ are restricted to take on discrete values~\cite{sonner2021thouless} $\tau \in \bZ$. When we rescale the values of $\{\theta_m\}$ to set the mean level spacing $\mu$ to unity,  this condition reads $\tau \in \mu \bZ$. On the other hand, for the DOG expression in \cref{eq:DOG_theta} to be well defined, $x$ can only take values on a circle and thus $\chi(x,N)$ has to be a periodic function of $x$ with period $2 \pi$ ($2 \pi/\mu$ when levels are rescaled to set mean level spacing to 1). Since we performed our analysis by selecting a relatively small subset of levels from the full spectrum, large values of $\theta_m - \theta_n$ are suppressed. Therefore, the SFF and DOG for circle-valued levels $\{\theta_m\}$ can be described by the expressions obtained for real-valued levels $\{E_m \}$ shown in \cref{eq:SFF_Chaos,eq:SFF_Poisson,eq:chi_poisson1,eq:chi_RMT}.

	\subsection{Spin chain with a quasi-periodic local field}
	Models of quasi-periodic MBL have now been studied in a variety of contexts through the properties of their eigenstates,  transport, and operator dynamics~\cite{iyer2013many,li2015many,Li2016,khemani2017two,lev2017transport,Setiawan-2017,Chandran-2017,vznidarivc2018interaction,Xu-2019,Yoo-2020,agrawal2020universality}.
	We consider the interacting spin chain system with a quasiperiodic magnetic field (QP) \cite{khemani2017two}, defined as:
	
	\begin{align}
		\label{eq:H_QP}
		H = J \sum_{i=1}^{L-1} (S_{i}^{x}S_{i+1}^{x}+S_{i}^{y}S_{i+1}^{y}) + J_{z} \sum_{i=1}^{L-1} S_{i}^{z}S_{i+1}^{z} + W\sum_{i=1}^{L} \cos(2\pi k i+ \phi)S_{i}^{z}  \nonumber\\   +J^{'} \sum_{i=1}^{L-2} \left(S_{i}^{x}S_{i+2}^{x}+S_{i}^{y}S_{i+2}^{y}\right)  
	\end{align}

	Here, $\phi \in [-\pi,\pi)$ is a site-independent phase offset used to generate an ensemble, $k=(\sqrt{5}-1)/2$ is an irrational number, and we set $J=J^{'}=J_{z}=1$ for numerical computation. With $\langle r \rangle $, the critical disorder is estimated to be around $W_{c} \approx 4.3$ \cite{khemani2017two}. When $W>W_c$, the system is in the MBL phase and when $W<W_c$ it is in the ergodic phase. This model also has spin-rotation $U(1)$ symmetry, and we consider the spectrum of the zero magnetization sector for numerical computations. 
	\begin{figure*}[!tbp]
		\centering
		\includegraphics[scale=0.353]{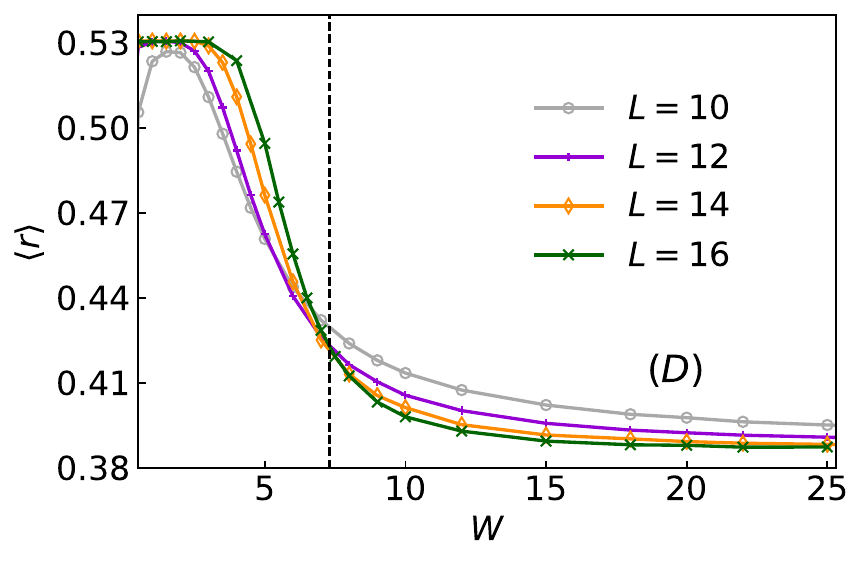}
		\includegraphics[scale=0.353]{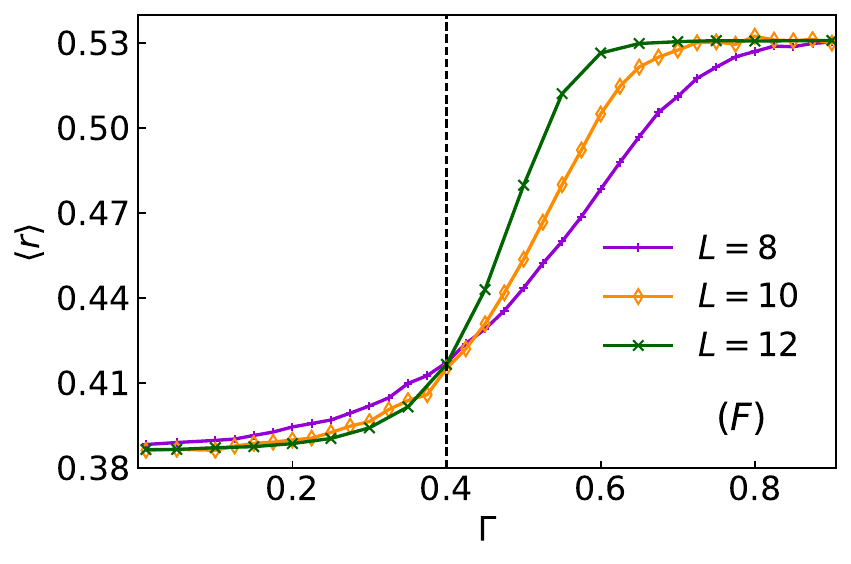}
		\includegraphics[scale=0.353]{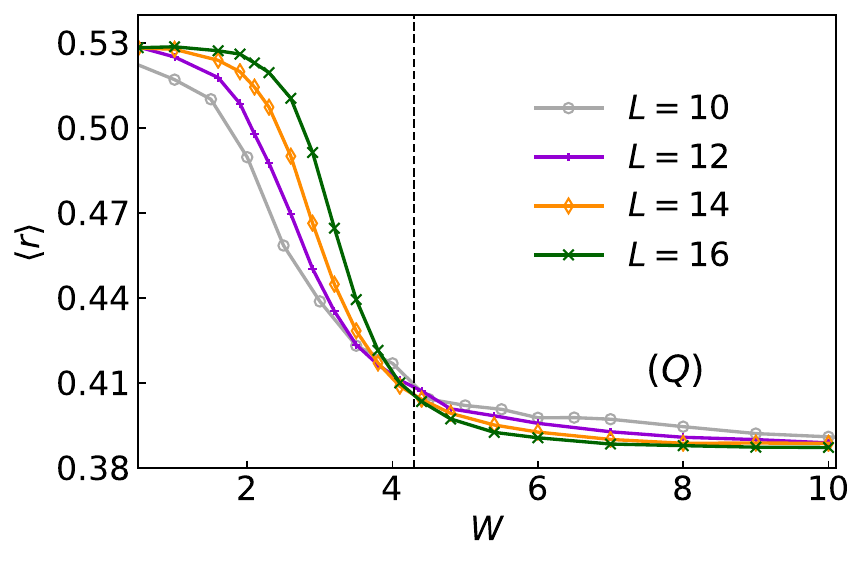}
		\includegraphics[scale=0.353]{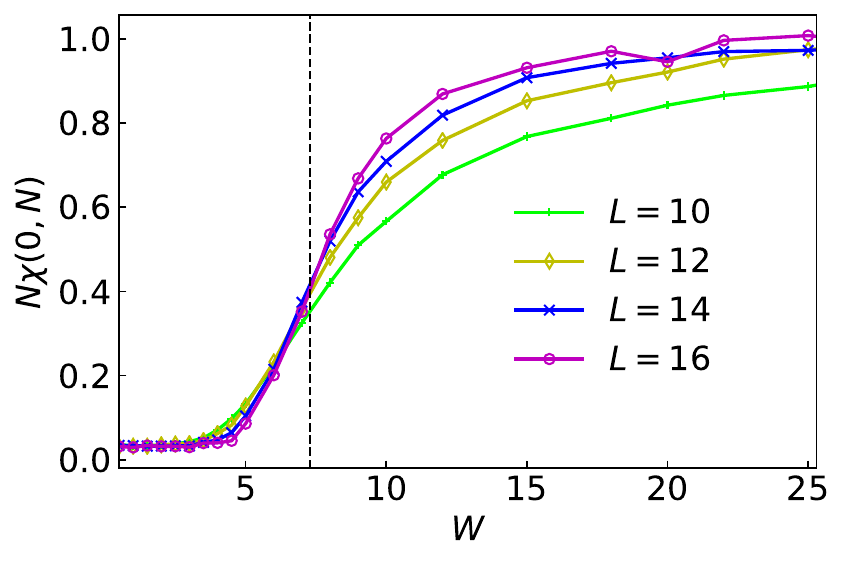}
		\includegraphics[scale=0.353]{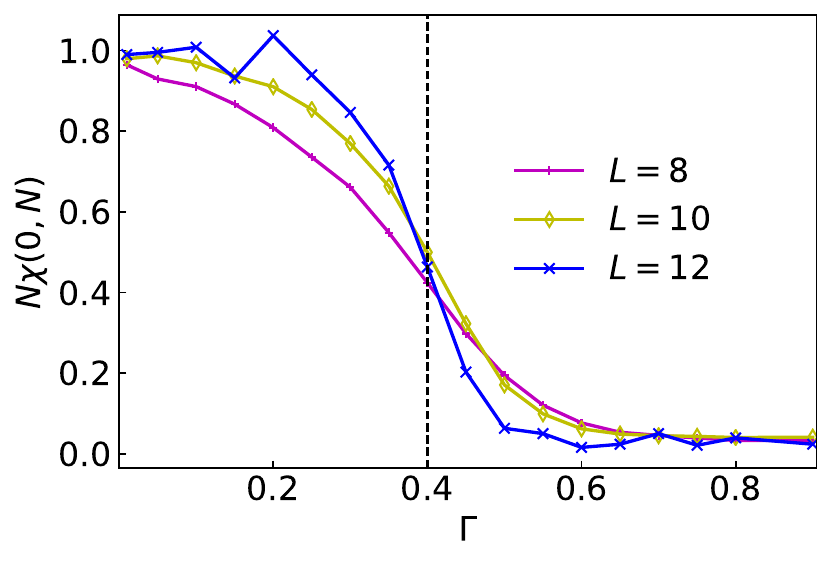}
		\includegraphics[scale=0.353]{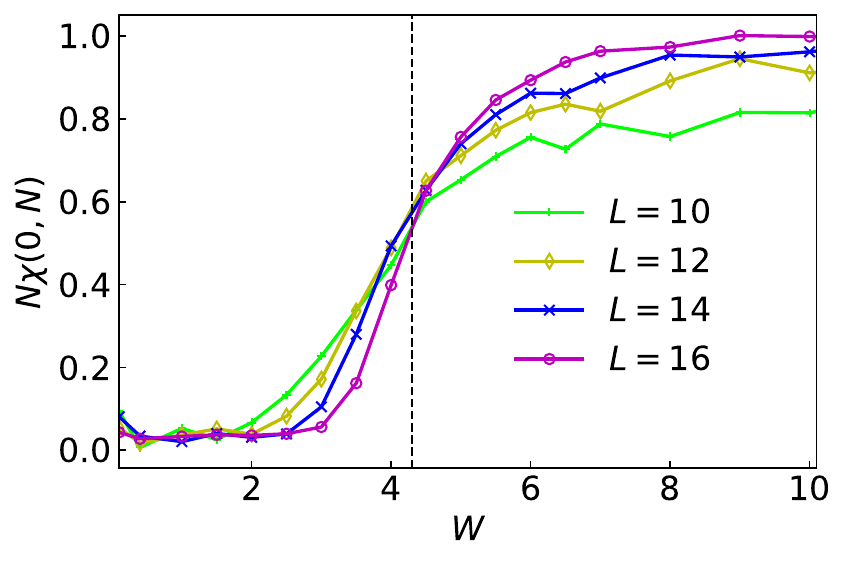}
		\caption {(Top row) Plots of the adjacent gap ratio $\langle r \rangle$ and (Bottom row) rescaled density of zero gaps ($N \chi(0,N)$) across various strengths and sizes of the system used to locate the approximate strength of the critical disorder (dashed line) where the curves for different sizes of the system cross for (from left to right) the disordered spin chain (D), Floquet spin chain (F) and quasi-periodic spin chain (Q) defined in \cref{eq:Hamiltonian,eq:Floquet unitary,eq:H_QP} respectively. The estimate of the critical disorder computed from $\moy{r}$ and $N \chi(0,N)$ agrees for all models. The full spectrum used for the plots in the top row and $N=20$ levels drawn from the middle of the spectrum used for plots in the bottom row.}
		\label{fig:r_chi0}
	\end{figure*}
	\subsection{Summary of results}
	The SFF for all three models [\cref{eq:Hamiltonian,eq:Floquet unitary,eq:H_QP}] is shown in the three columns of \cref{fig:SFF_numerics} from left to right, labeled (D), (F) and (Q) respectively.
	The SFF for disordered spin chain model [\cref{eq:Hamiltonian}] was studied in \cite{PrakashPixleyKulkarni_MBL_SFF_2021_PhysRevResearch.3.L012019} where it was verified that the numerical results match the analytical expressions very well. We reproduce the plots in \cref{fig:SFF_numerics} (left column) for completeness.  For small values of $N$, we see in \cref{fig:SFF_numerics}  that the SFF expressions \cref{eq:SFF_Poisson,eq:SFF_Chaos} match the data well deep in the MBL and ergodic phases, respectively. The power-law scaling form deep in the MBL phase is also verified in \cref{fig:SFF_numerics}. The DOGs for all three models are shown in \cref{fig:DOG}. We see that deep in the MBL and ergodic phases, the analytical expressions \cref{eq:chi_poisson1,eq:chi_RMT} describe the DOG well. Finally, in \cref{fig:r_chi0} (bottom row), we show the utility of the density of zero gaps $N \chi(0,N)$
to estimate the critical disorder strength corresponding to the MBL to ergodic transition using the location where the curves for various sizes of the system cross. Moreover, we see that it agrees well with the same result obtained from the adjacent gap ratio in \cref{fig:r_chi0} (top row).
	
\begin{figure}[!hbt]
		\centering
		\includegraphics[scale=0.84]{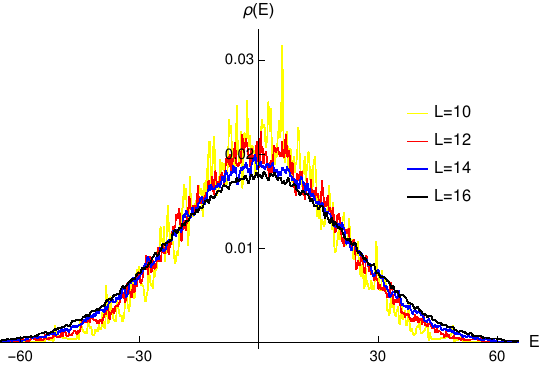}
		\includegraphics[scale=0.84]{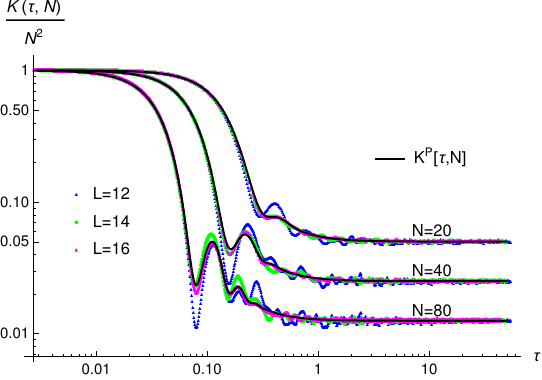}
		\caption{ {  Left: Persistent oscillations seen in the density of states of the full $S_z=0$ spectrum. Right: SFF [\cref{eq:SFF}] plotted for levels drawn form the middle of the $S_z=0$ sector of the spin chain with quasi-periodic local field [\cref{eq:H_QP}] deep in the MBL phase ($W=15$). These oscillations remain unchanged with increased disorder  averaging, but decrease in amplitude with increase in system size.   \label{fig:QP_oscillations}}}
	\end{figure}

	For the spin chain with a quasiperiodic local field [\cref{eq:H_QP}], we observe certain finite-size oscillatory features in the SFF, DOG and also in the density of states (DOS) as shown in \cref{fig:QP_oscillations} which are not removed by increasing the number of samples the data are averaged over. However, the amplitude of oscillations reduces with increasing system size as seen in \cref{fig:QP_oscillations}. We postulate that these finite-size oscillations have an origin in the quasiperiodic nature of on-site detuning. 
	
	\section{On the universality and robustness of spectral signatures in the MBL phase}
	\label{sec:Origin}
	Spectral signatures in the ergodic phase, for example, the linear ramp in the SFF, are considered to be universal features arising from the underlying level correlations in the system. This means that the ramp survives global deformations of the spectrum, such as unfolding. MBL, on the other hand, is often characterized by the absence of RMT features and level repulsions~\cite{vsuntajs2020ergodicity}. It is desirable to identify spectral signatures unique to the MBL phase beyond the absence of ergodic signatures. The results in the preceding sections are an attempt to isolate such signatures. The analytical expressions for the SFF and DOG, especially the power-law scaling form with a fixed exponent shown in \cref{eq:KPoisson_universal} was studied in Ref.~\cite{PrakashPixleyKulkarni_MBL_SFF_2021_PhysRevResearch.3.L012019}. This was shown to be a robust feature of the MBL spectra and can be thought of as a universal signature that can help to identify the MBL phase and distinguish it from the ergodic phase. However, a natural question is what the origin of these signatures in the MBL phase is and how robust they are. Since it is known, by the very definition, that uncorrelated Poisson numbers cannot have intrinsic correlations, their spectral signatures must have their origin in global effects. 
	
	In this section, we clarify robustness of universal MBL signatures to smooth changes in global density of states using deformations of various kinds to the spectrum and its probes. First, we study the effects of sharp features in the DOS of Poisson numbers generated as shown in \cref{sec:SFF_Poisson} and we study the robustness of the SFF as they are smoothed. We find that key features of the SFF, such as power-law scaling ~\cref{eq:KPoisson_universal} survive such changes. Next, we modify the expression for the SFF formula following the prescription in Ref.~\cite{vsuntajs2020ergodicity} and find that while the power-law scaling form survives, the coefficient is renormalized and eventually disappears. Finally, we compute the SFF after unfolding the spectrum, where we find that the power-law scaling form is eliminated. However, we find that it is again recovered if the SFF is computed from a subset of the levels in the unfolded spectrum. We conclude that the power-law scaling form has its origin in the global features of Poisson numbers but not directly from the density of states. They are nevertheless useful to distinguish MBL from ergodic systems in finite system size simulations, especially when probed at energy scales where the latter is dominated by intrinsic level repulsions. 
	\begin{figure*}[!hbt]
		\centering
		\includegraphics[width=0.32\textwidth]{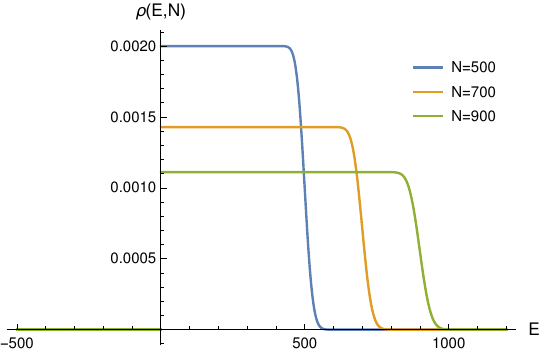} 
		\includegraphics[width=0.32\textwidth]{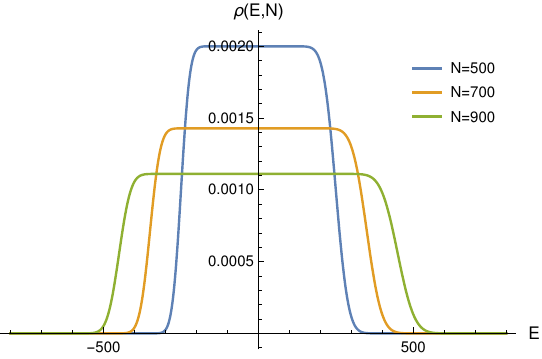}
		\includegraphics[width=0.32\textwidth]{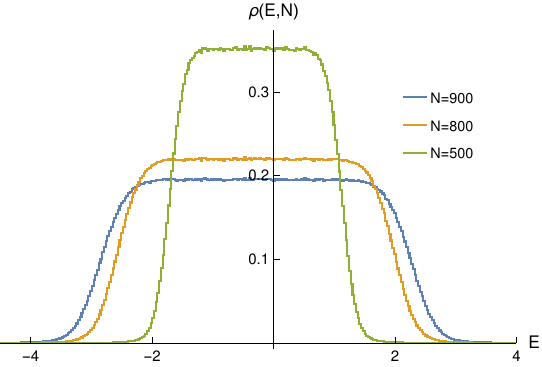}
		\includegraphics[width=0.32\textwidth]{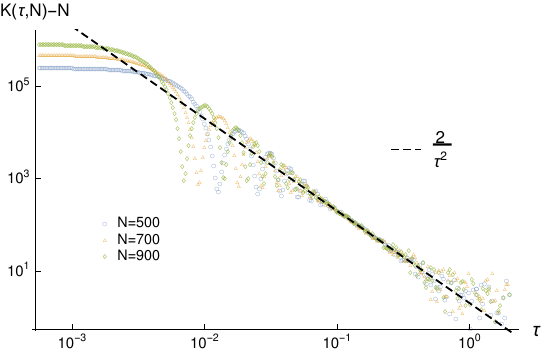}
		\includegraphics[width=0.32\textwidth]{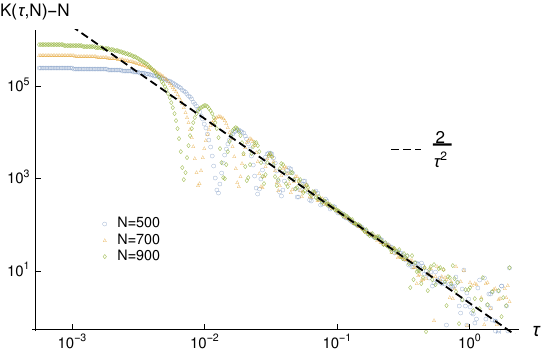}
		\includegraphics[width=0.32\textwidth]{H.pdf}
		\caption{ DOS (top row) and the reduced SFF (bottom row) for Poisson numbers generated as described in \cref{eq:built_spectrum} (left column), Poisson numbers generated in \cref{eq:built_spectrum} with random offset to soften the DOS edges, shown in \cref{eq:smoothenedDOS} (middle column) and levels chosen from the disordered Hamiltonian \cref{eq:Hamiltonian} ($L=18$) deep in the MBL phase ($W=25$).}
		\label{fig:Poisson_XXZ_DOS_SFF}
	\end{figure*}
	\subsection{Deforming the global density of states}
	
	Using the probability density in \cref{eq:Poisson_2pt}, we can easily compute the density of states for Poisson numbers generated as detailed in \cref{sec:SFF_Poisson} explicitly as~\cite{PrakashPixleyKulkarni_MBL_SFF_2021_PhysRevResearch.3.L012019} 
	\begin{align}
		\rho(E,N) &= \frac{1}{N} \sum_{k=1}^N \moy{\delta(E - E_k)} = \frac{\Gamma\left(N,E/ \mu \right)}{\mu~N!}, \label{eq:Poisson_DOS}
	\end{align}
	where, $\Gamma(N,x)$ is the incomplete Gamma function defined (for integer $N$) as
	\begin{align}
		\Gamma(N,x) = \int_x^\infty dt~t^{N-1} e^{-t} = e^{-x}~ (N-1)!  \sum_{k=0}^N \frac{x^k}{k!}.
	\end{align}
	Since the energies are chosen to be positive definite, the ensemble-averaged DOS has a sharp edge at $E=0$ as shown in the top left column of \cref{fig:Poisson_XXZ_DOS_SFF}. One might postulate that the power-law scaling form is an artifact of Fourier transforming this sharp edge. However, this is not the case, as easily seen from the formula for the SFF~\cref{eq:SFF} which only takes into account the differences in levels and would not change if the spectrum were modified to smooth out the sharp edge. For instance, let us shift all Poisson numbers by the same random offset parameter $E_{0}$, drawn from a Gaussian distribution $p_0(x)$ with mean $N/2$ and variance $N$,
	\begin{equation}
		p_0(x) =\frac{1}{\sqrt{2 \pi N}} \exp{\left(-\frac{(x-N/2)^2}{2N}\right)}.
	\end{equation}
	Assuming  $m>n$, the new joint two-point distribution $\tilde{P}(E_n,n;E_m,m)$ is 
	\begin{align}
		\tilde{P}(E_n,n;E_m,m) = \tilde{p}(E_n,n)~ p(E_m-E_n,m-n), \label{eq:smoothenedDOS}
	\end{align}
	where, 
	\begin{align}
		p(E,k) &=   \frac{e^{-\frac{E}{\mu} }}{\mu (k-1)!} \left(\frac{E}{\mu}\right)^{k-1},\\
		\tilde{p}(E,k) &=\int_0^\infty dx~ p(x,k)~ p_0(E-x).
	\end{align}
The form of the new density of states,	$\tilde{\rho}(E,N)$ can be determined as
	\begin{align}
		\tilde{\rho}(E,N) &= \int_0^\infty dx ~\rho(x,N) ~p_0(E-x) \label{eq:Modified DOS} 
	\end{align}
	where, $\rho(x,N)$ is the original DOS given in \cref{eq:Poisson_DOS}. By numerical integration \cref{eq:Modified DOS}, we can see in \cref{fig:Poisson_XXZ_DOS_SFF} (top row, middle column) that the DOS is smoothed out. However, as mentioned before, the calculation of the SFF \cref{eq:SFF} only involves the distribution of energy differences $E_m - E_n$ and therefore is completely independent of the change in DOS introduced in \cref{eq:smoothenedDOS} and leaves both analytical expressions of \cref{eq:SFF_Poisson,eq:KPoisson_universal} unchanged (see the bottom row, middle column of \cref{fig:Poisson_XXZ_DOS_SFF}) 
	
	In fact, this was already verified directly in \cref{fig:SFF_numerics} where plots there were computed by selecting levels from the middle of the spectrum. As seen in \cref{fig:Poisson_XXZ_DOS_SFF} (top right), the density of states for these levels has no sharp edges, and we clearly see that the power-law scaling is robust. To end this section, we remark that the formula for the connected SFF shown in \cref{eq:SFFC} does have a dependence on DOS. However, the effect of smoothing the DOS as \cref{eq:smoothenedDOS}  is to enhance (double) the coefficient of the power-law scaling form as
	\begin{equation}
		K_c(\tau,N) - N = \frac{1}{(\mu \tau)^2 } \mapsto  \frac{2}{(\mu \tau)^2 }.
	\end{equation}
	See \ref{app:ConnectedSFF} for more details on this. 
	\subsection{Filtering the SFF formula}
\begin{figure}[!h]
		\centering
		\includegraphics[scale=0.86]{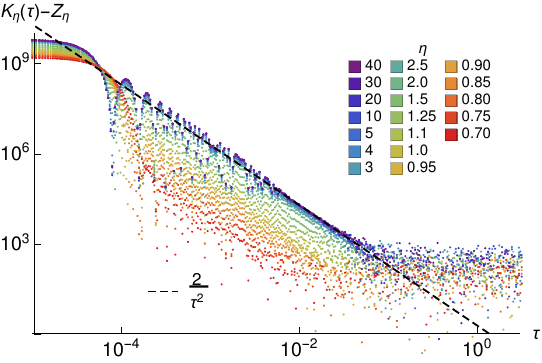}
		\includegraphics[scale=0.56]{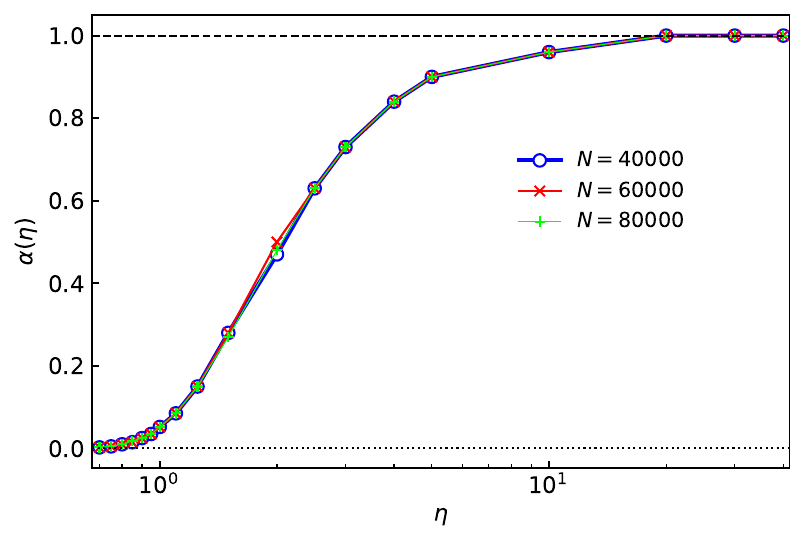}
		\caption{ {  Left: The reduced filtered SFF \cref{eq:filter_SFF} for $N=$ 80,000 Poisson numbers generated as shown in \cref{eq:built_spectrum} averaged over 50,000 disorder samples. Right:  Plot of modified coefficient of the power-law scaling $\alpha$, defined in \cref{eq:filter_reduced_SFF} for various $\eta$. It is clear that $\alpha(\eta)$ has no dependence on $N$. }}
		\label{fig:filtering_SFF}
	\end{figure}

	We now modify the expression for SFF \cref{eq:SFF} using the prescription of Ref. \cite{vsuntajs2020quantum} by introducing weights $\zeta(E)$ associated with the eigenvalue $E$ to define the filtered SFF expression
	\begin{equation}
		\label{eq:filter_SFF}
		K_{\eta}(\tau)=\langle \sum_{m,n=1}^{Nr}\zeta(E_{m})\zeta(E_{n})e^{i \tau(E_{m}-E_{n}} \rangle.
	\end{equation}
	Above, $\zeta(E)$ is a Gaussian filter function defined as
	\begin{align}
		\label{eq:gaussian}
		\zeta(E)=\exp\bigg[\frac{-(E-\bar{E})^2}{2(\Gamma \eta)^2}\bigg] ,
	\end{align} 
	where, $\bar{E}$ and $\Gamma$ are set to the mean and variance of levels from each disorder realization. In the limit $\tau \rightarrow \infty$, $K_{\eta}(\tau)$ saturates to $Z_{\eta}$, given by the expression
	\begin{equation}
		Z_{\eta}=\langle \sum_{m=1}^{Nr}|\zeta(E_{m})|^2 \rangle.   \label{eq:filter_saturation}  
	\end{equation}
	The parameter $\eta$ controls the width of the filter. As seen in \cref{fig:filtering_SFF}, as $\eta$ is reduced, the filtered SFF preserves the power-law scaling form but renormalizes the coefficient as
	\begin{equation}
		\label{eq:filter_reduced_SFF}
		K_{\eta}(\tau) - Z_{\eta} = \frac{2}{(\mu \tau)^2}  \mapsto \alpha(\eta)  \frac{2}{(\mu \tau)^2},
	\end{equation}
	where, $\alpha(\eta)$ is a function that depends on the filter parameter $\eta$ but not on the number of levels $N$ as shown in \cref{fig:filtering_SFF}.
	A spectral signature that does not have its origin in global features would be expected to survive in the limit when we take $N$ to be large and $\eta$ to be small. This is true for the linear ramp in the ergodic phase~\cite{vsuntajs2020quantum}. However, since $\alpha$ has no dependence on the number of levels, in the limit of large $N$ and small $\eta$, it is expected that the power law scaling will vanish. This is consistent with the assertion that the power-law scaling has its origin in global features. 
	
	\subsection{Unfolding the spectrum}
\begin{figure}[!h]
		\centering
		\includegraphics[scale=0.85]{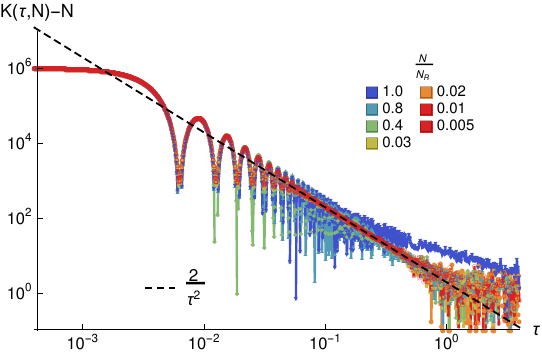}
		\includegraphics[scale=0.55]{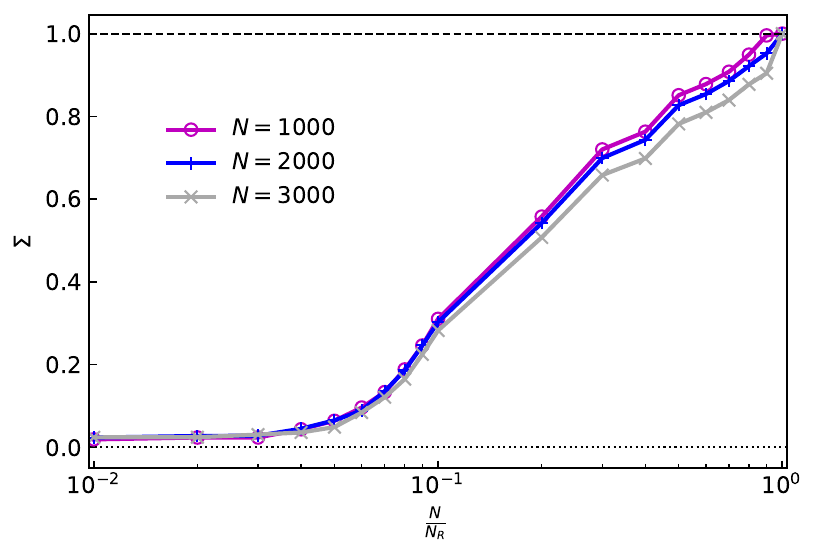}
		\caption{{  Left: Reduced SFF for $N=1000$ fixed levels selected from unfolded $N_R$ Poisson numbers generated as shown in \cref{eq:built_spectrum} averaged over 400,000 disorder samples. Right: RMS deviation of the SFF $\Sigma$ from the power law scaling form, defined in \cref{eq:std_deviation} calculated with the data points in the window $\tau \in [\frac{3}{\sqrt{N}},0.2]$. The power-law scaling form is absent when $\frac{N}{N_R} \sim 1$ and is recovered at the limit $\frac{N}{N_R} \sim 0$.}}
		\label{fig:CSFF_poisson_unfolded}
	\end{figure}

	In this section, we discuss the effect of unfolding the spectrum on the SFF. We begin with Poisson numbers generated as mentioned in \cref{sec:SFF_Poisson} whose SFF matches the analytical expression in \cref{eq:SFF_Poisson} and exhibits power law scaling \cref{eq:KPoisson_universal} at intermediate $\tau$ values. We now unfold~\cite{french1971some,mayer1997,bruus1997,guhr1999} the same levels using the usual polynomial method~\footnote{We first find the cumulative density of the ordered original spectrum $\{E_n\}$ i.e. $I(E)=\sum_n \Theta(E-E_n)$, where $\Theta(x)$ is the Heaviside step function, and then find a  smooth polynomial function $\tilde{I}(E)$ to fit the data of $I(E)$. With that we obtain the unfolded spectrum as $\tilde{E}_n=\tilde{I}(E_n)$.  For our numerical fits, we used a $15^{th}$ order polynomial.}. Although the density of states was uniform before and after unfolding~\cite{riser2020nonperturbative}, we see in \cref{fig:CSFF_poisson_unfolded} that the SFF is no longer described by \cref{eq:SFF_Poisson} and does not exhibit power law scaling \cref{eq:KPoisson_universal}. Now, if we compute the SFF with decreasingly smaller fractions of the unfolded spectrum, the analytical form as well as the power law scaling is recovered as shown in \cref{fig:CSFF_poisson_unfolded} (top). To quantify this, we compute the root-mean-square (RMS) deviation of the SFF from the power law scaling form, $\Sigma$, defined as 
	\begin{equation}
		\label{eq:std_deviation}
		\Sigma^2 =\Big\langle \Big( K(\tau,N)-N-\frac{2}{\tau^2} \Big)^2 \Big\rangle \\- \Big( \Big\langle K(\tau,N)-N-\frac{2}{\tau^2} \Big\rangle \Big)^2,
	\end{equation}
	where $\langle \dots \rangle$ represents the average taken over data points falling in the domain $\tau \in [\frac{3}{\sqrt{N}},0.2]$ where the power law scaling form is expected to be clearest when present. As seen in \cref{fig:CSFF_poisson_unfolded} (bottom), for $N/N_R \rightarrow 1$, $\Sigma$ is large, indicating that the SFF deviates significantly from the power-law scaling form. As $N/N_R \rightarrow 0$, we find that $\Sigma$ vanishes, indicating that the power-law scaling form is recovered.

	We conclude this section with a few comments. Unlike the linear ramp of the ergodic SFF which is unaffected by unfolding, the power-law scaling form of the SFF in the MBL phase is a global feature. The long-ranged spectral signatures for the MBL phase, such as power-law scaling, arise from the global features of ordered, uncorrelated levels. This is related to the emergence of the Poisson distribution in the distribution of spectral gaps starting with an ensemble of random numbers drawn from any independent identical distribution after ordering (see the supplementary material of Ref. \cite{PrakashPixleyKulkarni_MBL_SFF_2021_PhysRevResearch.3.L012019} for a discussion on this).  In conclusion, we believe that our results suggest that using spectral signatures that have their origin in both intrinsic and global details is more useful in characterizing eigenstate phases as well as integrability (and its breaking) on a finite number of levels as compared to those that are retained by unfolding and similar procedures and depend on intrinsic correlations only.
	For example, the analysis presented here is quite useful to characterize the spectral correlations in the middle part of the many-body spectrum at finite sizes, as conventionally considered in numerical studies of MBL.

	\begin{figure}[!hbt]
		\centering
		\includegraphics[scale=0.56]{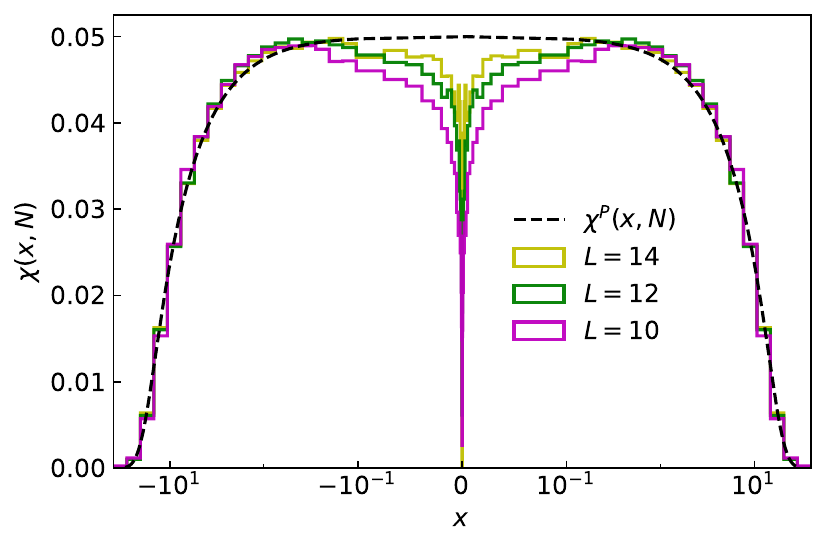}
		\includegraphics[scale=0.56]{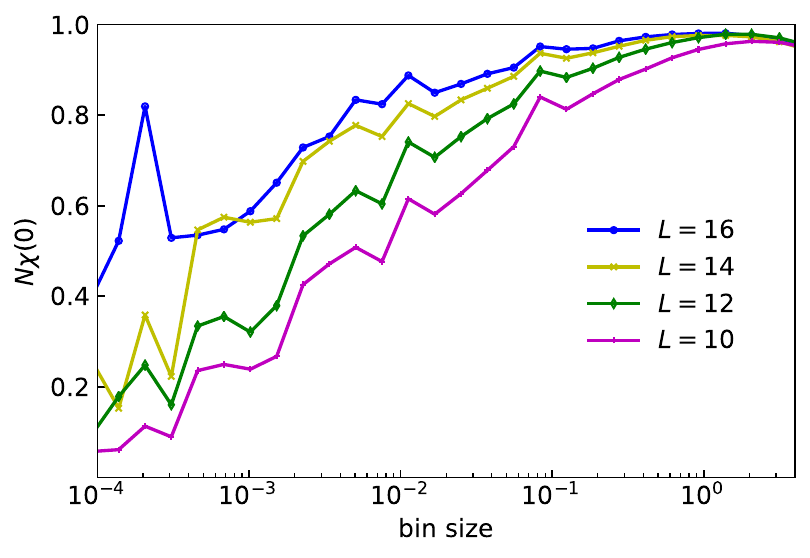}
		\caption{{  Left: DOG [\cref{eq:DOG}] for the disordered spin chain model \cref{eq:Hamiltonian} in the MBL phase with disorder strength $W=15$ computed with $N=20$ energy levels drawn from the middle of the spectrum plotted on a log scale to capture the dip at the origin. The DOG approaches the form predicted by Poisson numbers~\cref{eq:chi_poisson1} with increase in system size. Right: Rescaled density of zero gaps ($N \chi(0,N)$) versus binning size. This vanishes for small bin and system size and approaches the value of $1$ predicted using Poisson numbers with increase in bin and system size.}} 
		\label{fig:chi_vs_bin}
	\end{figure}
	
	\section{Discussion}
	\label{sec:Other}
	In this final section, we discuss our findings relative to other related works of long-ranged spectral correlations in the MBL phase.
	
	\subsection{The effect of many-body resonances in the MBL phase}
	\label{sec:level_repulsion}
	
	In this paper, we have used Poisson numbers to derive analytical expressions for the SFF and DOG in the MBL phase, which assumes that the levels have zero intrinsic correlations. 
	While this should be a good description deep in the MBL phase, as we approach the transition, however, many-body resonances will begin to proliferate~\cite{DeRoekHuveneers_bubble_PhysRevB.95.155129,DeRoeckImbrie_2017many}, an effect that we have neglected in our treatment of the SFF. These resonances can lead to weak but nonzero level repulsions in finite-size systems that have been shown to have a nonvanishing value of $\chi(0,N)$~\cite{GarrattRoyChalker_MBL_SFF_DOGPhysRevB.104.184203}  if the energy resolution is chosen appropriately (and we note that this effect vanishes in the thermodynamic limit). Indeed, these effects are also observable in our finite-size simulations and in the appropriate regimes connects to our results from the previous sections as we now demonstrate. 
	
	The expression of DOG for Poisson  numbers, $\chi^P(x,N)$ shown in \cref{eq:chi_poisson1} shows a peak for zero-gaps, $\chi(0,N)$ as shown in \cref{fig:DOG_poisson_RMT} which is confirmed in numerical studies of $\chi(x,N)$ in \cref{sec:Numerical}. To see the vanishing of $\chi(0,N)$ predicted in \cite{GarrattRoyChalker_MBL_SFF_DOGPhysRevB.104.184203}, we consider the disordered spin chain model shown in \cref{eq:Hamiltonian}. \Cref{fig:chi_vs_bin} shows that with careful binning on a log-scale, $\chi(x,N)$ deviates from $\chi^P(x,N)$ in a range $\delta_L$ and vanishes in the limit $x \rightarrow 0$. With increasing system size, we also see that $\delta_L$ reduces and the DOG increasingly agrees with \cref{eq:chi_poisson1}. If we increase the size of the histogram bins to be larger than $\delta_L$, as it was done in \cref{sec:Numerical}, the finite-size vanishing of $\chi(0,N)$ in the MBL is not observed, as seen in \cref{fig:DOG} and we find $N\chi(0,N) \rightarrow 1$ as expected from \cref{eq:chi_poisson1}. More details on the robustness of $\chi(x,N)$ for binning sizes larger than $\delta_L$ are shown in \ref{app:binning}.
	
	In summary, if we choose a bin size larger than $\delta_L$, \cref{eq:chi_poisson1} reproduces the form of $\chi(x,N)$ accurately. As we increase system size $L$ and disorder strength, we expect $\delta_L$ to reduce, and the expression of \cref{eq:chi_poisson1} is increasingly accurate for various binning sizes. 
	
	\subsection{Symmetry breaking picture for delocalization}
	\label{sec:SSB_picture}
	\Cref{{fig:DOG_poisson_RMT}} shows the DOG for Poisson numbers and random matrices. It is clear that the DOG distinguishes the two cases very well. While the DOG for random matrices is characterized by a deficit in zero gaps, the Poisson DOG is characterized by a clustering at zero gaps. From its definition in \cref{eq:DOG}, $\chi(x,N)$ has a reflection symmetry
	\begin{equation}
		\chi(-x,N) = \chi(x,N). \label{eq:chi_symmetry}
	\end{equation}
	The value of spectral gaps $x^*$ that maximizes $\chi$ serves as a proxy for level repulsions. From \cref{fig:DOG_poisson_RMT} it is clear that for RMT, $x^* \neq 0$ and for Poisson numbers, $x^* \rightarrow 0$. Thus, the $x \mapsto -x$ symmetry of $\chi(x,N)$ leaves the value of $x^*$ invariant in the MBL phase but not in the ergodic phase. This is reminiscent of the symmetry-breaking transition in the Ising model if we make an analogy between$-\chi(x,N)$ with the Landau free-energy potential of the Ising model and $x^*$ the location of the potential minima with the magnetic order parameter. 
	This motivates the possibility of tracking the MBL-to-ergodic transition via a symmetry-breaking framework.
	We remark that a symmetry-breaking framework to describe the ergodic-MBL transition was recently discussed in Ref.~\cite{GarrattChalker_MBLSSB_PhysRevLett.127.026802} where the ergodic phase also corresponded to the symmetry-breaking phase. However, the symmetry they consider is more complex, and it is unclear if and how it is connected to the $\ztwo$ symmetry of the DOG described above.

	\section{Summary and Outlook}
	
	In this paper, we have studied long-range spectral probes and their use in characterizing many-body localization and ergodic phases and the transition between them. We obtain analytical forms for these probes and numerically verify their validity and utility using a variety of spin chain models to find excellent agreement. We also discussed the nature of universality and robustness of these spectral signatures and briefly commented on related work. {  Recent demonstration of large-scale Hamiltonians involving a large number of qubits (or spin sites) in quantum devices is indeed a promising avenue to potentially explore these quantities. The robust nature of these quantities that we demonstrated in this work indicates that such quantities can potentially be observed despite experimental imperfections that are often inevitable in large scale systems.} In future work, it would be interesting to further explore the utility of long-range spectral probes to shed light on the nature of the transition between many-body localization and ergodic phases, as well as investigate whether the symmetry breaking picture of \cref{sec:SSB_picture} can help produce an effective theory for the transition.

	\section*{Acknowledgements}
	
	The authors thank David Huse and Sthitadhi Roy for useful discussions. For most of this work, A.P. was supported by a grant from the Simons Foundation (677895, R.G.) through the ICTS-Simons prize postdoctoral fellowship. He is currently funded by the European Research Council under the European Union Horizon 2020 Research and Innovation Programme, Grant Agreement No. 804213-TMCS. M.K would like to acknowledge support from project 6004-1 of the Indo-French Centre for the Promotion of Advanced Research (IFCPAR), Ramanujan Fellowship (SB/S2/RJN-114/2016), SERB Early Career Research Award (ECR/2018/002085) and SERB Matrics Grant (MTR/2019/001101) from the Science and Engineering Research Board (SERB), Department of Science and Technology, Government of India. MH, AP, and MK acknowledge support from the Department of Atomic Energy, Government of India, under Project No. RTI4001. J.H.P. is partially supported by NSF Career Grant No. DMR- 1941569 and the Alfred P. Sloan Foundation through a Sloan Research Fellowship.
	J.H.P. acknowledges the Aspen Centre for Physics, where some of this work was performed, which is supported by National Science Foundation grant PHY-1607611.
	M.K and J.H.P acknowledge the support from the Rutgers Global International Collaborative Research Grant. M.K. acknowledges support from the Infosys Foundation International Exchange Program at ICTS.


	\appendix
	
	\section{Computing the statistics of $r_i$} \label{app:r}
	
	The distribution of $r_i$, $P(r_i)$ can be computed using the local distributions of consecutive gaps $P(\delta_i,\delta_{i+1})$ as follows
	\begin{equation}
		P(r_i) = \int_0^\infty d \delta_i \int_0^\infty d \delta_{i+1}~ P(\delta_i,\delta_{i+1}) \\ \bigg[\delta\left(r_i - \frac{\delta_i}{\delta_{i+1}}\right) \Theta \left(\delta_{i+1}-\delta_i\right) + \\ \delta\left(r_i - \frac{\delta_{i+1}}{\delta_{i}}\right) \Theta \left(\delta_i-\delta_{i+1}\right)\bigg] 
	\end{equation}
	where, $\delta(x)$ and $\Theta(x)$ are the Dirac delta function and Heaviside step functions respectively. A few simplifications leads to
	\begin{equation}
		P(r_i) = 
		\int_0^\infty d \delta_i P(\delta_i,r \delta_i) \delta_i \\+\int_0^\infty d \delta_{i+1} P(r\delta_{i+1}, \delta_{i+1}) \delta_{i+1}
	\end{equation}
	$P(r_i) =0$ unless $r_i \in (0,1)$. An expression for $P(r)$ can be obtained by assuming that consecutive gaps are uncorrelated and identically distributed thereby approximating the local distributions of consecutive gaps $P(\delta_i,\delta_{i+1})$ as
	\begin{equation}
		P(\delta_i,\delta_{i+1}) \approx P(\delta_i) P(\delta_{i+1}) \label{eq:uncorrelated nn level}
	\end{equation}
	where,  $P(\delta_i)$ is the distribution of the nearest neighbour gaps that we approximate as independent of the index $i$. This reduces the equations for $P(r_i)$ to the following
	\begin{equation}
		P(r) = 2 \int_0^\infty d \delta P(\delta) P(r \delta) \delta \label{eq:P(r) nn approximation} 
	\end{equation}
	For the case of RMT ensembles, the distribution of nearest-neighbour level spacings can be written using the Wigner surmise~\cite{mehta2004random,Haake_QuantumChaosBook} as
	\begin{equation}
		P_\beta(\delta) = a_\beta \delta^\beta e^{-b_\beta \delta^2}.
	\end{equation}
	Where, $\beta$ indexes the ensemble and corresponds to GOE ($\beta = 1$), GUE ($\beta = 2$) or GSE ($\beta = 4$). $a_\beta$ and $b_\beta$ are ensemble-dependent constants. Using \cref{eq:P(r) nn approximation}, we can obtain the distribution of adjacent gap ratio as
	\begin{equation}
		P_\beta(r) = c_\beta \frac{r^\beta}{(1+r^2)^{\beta+1}},
		\label{eq:Pr_chaos}
	\end{equation}
	where $c_\beta$ is a normalization constant that depends on $\beta, a_\beta$ and $b_\beta$ 
	\begin{equation}
		c_\beta = \frac{\beta!~ a_\beta }{b_\beta^{\beta+1}} 
	\end{equation}
	that can be easily fixed by requiring unit normalization of $P(r)$ ($c_1 = 2^2$, $c_2 = \frac{2^5}{\pi}$, $c_4 = \frac{2^9}{3 \pi}$). Better approximations to the expression of $P(r)$ can be obtained by introducing corrections to \cref{eq:uncorrelated nn level} ~\cite{AtasBogomolnyGiraudRoux_adjacentgapratioFormula_PhysRevLett.110.084101}.
	
	For Poisson spectra on the other hand, substituting, the level spacings are exponentially distributed
	\begin{equation}
		P(\delta) = e^{-\delta}.
	\end{equation}
	Using \cref{eq:P(r) nn approximation}, we get~\cite{OganesyanHuse_2007_PhysRevB.75.155111}
	\begin{equation}
		P(r) = \frac{2}{(1+r)^2}.
		\label{eq:mbl_pr}
	\end{equation}
	Keeping track of the value of $\moy{r}$ can tell us if the system is chaotic or integrable/ many-body localized. If the system is in the MBL phase, using \cref{eq:mbl_pr} we get $\langle r \rangle = \int_{0}^{1} r P(r) dr =2\log 2-1 \approx 0.39$. On the other hand, if the system is in the ergodic phase, then 
	using the approximated \cref{eq:Pr_chaos} , $\langle r \rangle$ gives
	\begin{equation}
		\langle r \rangle = c_{\beta}\,\, \frac{_2F_1[1+\beta, 2+\beta,3+\beta,-1]}{2+\beta} 
		\label{eq:r_chaosF}
	\end{equation}
	where $_2F_1$ is the standard Gauss Hypergeometric function. This \cref{eq:r_chaosF} for GOE ($\beta=1$), GUE ($\beta=2$), GSE ($\beta=4$) gives us $0.57, 0.63, 0.70$ respectively. 
	This is reasonably close to what we expect by direct numerical simulation on random matrices $(0.53, 0.59, 0.67)$  
	and better approximations to the expression of $P(r)$ can be obtained by introducing corrections to \cref{eq:uncorrelated nn level} ~\cite{AtasBogomolnyGiraudRoux_adjacentgapratioFormula_PhysRevLett.110.084101} which will yield  $\langle r \rangle$ that is more accurate. 
	\begin{figure*}[!hbt]
		\centering
		\includegraphics[scale=0.35]{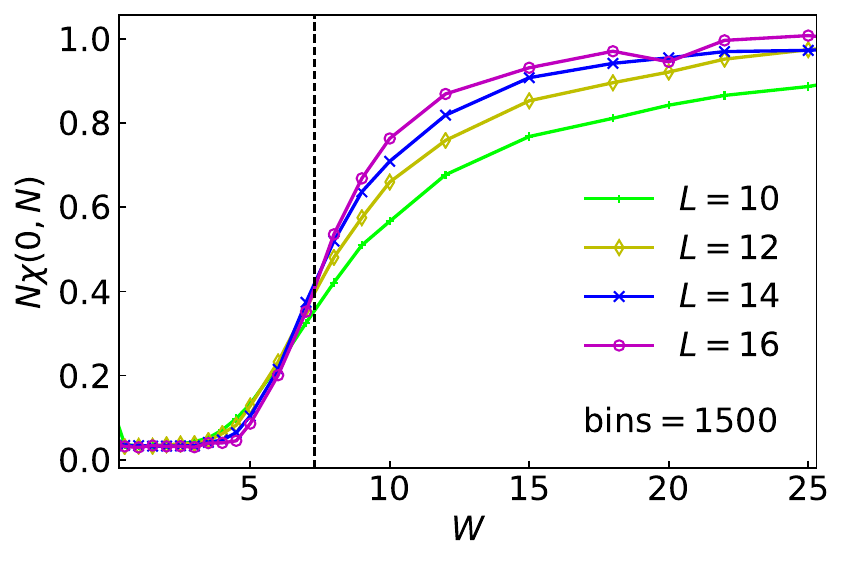}
		\includegraphics[scale=0.35]{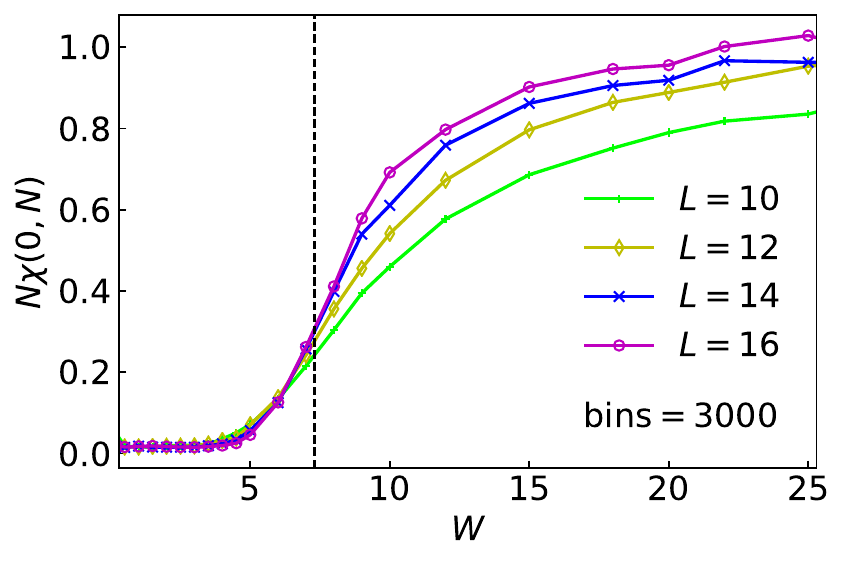}
		\includegraphics[scale=0.35]{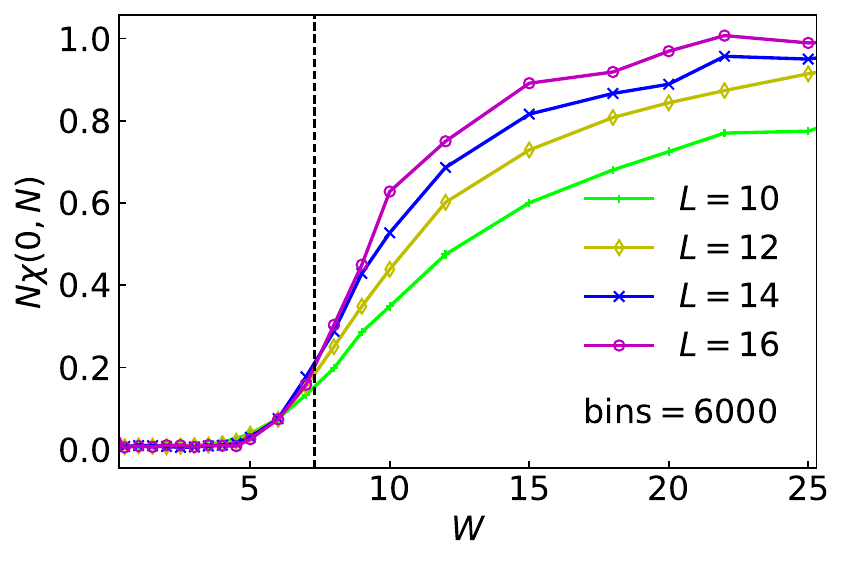}
		\caption{ Plots of $N \chi(0,N)$ for the disordered spin chain~\cref{eq:Hamiltonian} versus disorder strength $W$ computed using different number of bins to construct histograms. Data averaged over $50000, 30000, 20000, 5000$ disorder samples for system sizes $L=10, 12,\ 14,\ 16$ respectively. The estimate for critical disorder strength $W^*$ (dashed line) corresponding to the location where curves of different system sizes cross is unchanged with binning sizes. $N=20$ levels used from middle of spectrum of each disorder sample and scaled to set mean level spacing to unity. }
		\label{fig:chi0_bins_critical}
	\end{figure*}
	
	\section{DOS effects on the connected SFF for Poisson numbers}
	\label{app:ConnectedSFF}
Let us consider the Poisson numbers generated as shown in \cref{eq:built_spectrum} which has the 2-point probability distribution  $P(E_n,n;E_m,m)$ given by
	\begin{align}
		P(E_n,n;E_m,m) = p(E_n,n)~ p(E_m-E_n,m-n),
	\end{align}
	where $p(E_k,k)$ is the well known Poisson distribution
	\begin{equation}
		p(E_k,k) =  \begin{cases}
			\frac{e^{-\frac{E_k}{\mu} }}{\mu (k-1)!} \left(\frac{E_k}{\mu}\right)^{k-1}~E_k \ge 0 \\
			~~~~~~~~0 ~~~~~~~~~~~~~E_k <0
		\end{cases} 
	\end{equation}
	with $E_k>0$ and $k = 1,2,3,\ldots$. As mentioned in the main text, the SFF and CSFF can be computed to get the following expressions (see Ref.~\cite{riser2020nonperturbative} and the supplementary materials of Ref.~\cite{PrakashPixleyKulkarni_MBL_SFF_2021_PhysRevResearch.3.L012019} for the derivation)
	\begin{align}
		K(\tau,N) &= N + \frac{2}{(\mu\tau)^2}  - \frac{ (1+i \mu \tau)^{1-N} + (1-i \mu \tau)^{1-N}  }{(\mu\tau)^2}, \label{eq:SFF_Poisson1} \\
		K_c(\tau,N) 
		&= N + \frac{1}{(\mu\tau)^2} - \frac{(1+(\mu\tau)^2)^{-N}}{(\mu\tau)^2} \nonumber \\  &~~~~~- \frac{i}{  \mu \tau } \left[ (1 + i \mu \tau )^{-N} - (1 - i \mu \tau )^{-N}  \right] . \label{eq:SFFC_Poisson}
	\end{align}
	
	We now focus on the intermediate $\tau$ regime $\frac{1}{N}< \mu \tau < 1$ where the expressions reduce to
	\begin{align}
		K(\tau,N)-N &=  \frac{2}{(\mu \tau)^2} + \ldots \label{eq:KPoisson_universal1}\\
		K_c(\tau,N)-N &=  \frac{1}{(\mu \tau)^2} + \ldots \label{eq:KcPoisson_universal}
	\end{align}
	
	A noticeable feature in \cref{eq:KPoisson_universal1,eq:KcPoisson_universal} is that the SFF and the CSFF have different overall coefficients in the power-law scaling. The reason for this is in the nature of the density of states (DOS) for the Poisson numbers discussed in the main text - since the energies are chosen to be positive definite, the ensemble-averaged DOS has a sharp edge at $E=0$ as shown in \cref{fig:Poisson_XXZ_DOS_SFF}. We argued that the sharp edge in the DOS has no effect on the SFF. To understand the effect on the connected SFF, let us rewrite the expression for the SFF and CSFF in terms of the DOS. For this, let us introduce the quantity $Z(\tau,N)$ which relates the SFF $(K(\tau,N))$ and CSFF $(K_c(\tau,N))$ as follows
	\begin{align}
		K(\tau,N) &= K_c(\tau,N) + |Z(\tau,N)|^2, \label{eq:app_SFFCSFFZ}\\
		Z(\tau,N)&=\langle \sum_{m=1}^N e^{i \tau E_m} \rangle = N \int_{-\infty}^{\infty} dE~ e^{- \tau E} \rho(E,N).
	\end{align}
	This means that $Z(\tau,N)$ is related to the Fourier transform of the many-body density of states. It is important to note that the effect of a hard edge in the spectrum by shifting the lowest energy state to zero, will have a strong effect in the Fourier transform as follows
	\begin{align}
		Z(\tau,N) &= \frac{1}{\mu \tau }  + \mathcal{O}\left(\frac{1}{N}\right), \\
		|Z(\tau,N)|^2 &= \frac{1}{(\mu \tau)^2}  + \mathcal{O}\left(\frac{1}{N}\right), \label{eq:Z(tau,N)}
	\end{align}
	\Cref{eq:app_SFFCSFFZ} tells us that the difference in the coefficient of the power-law scaling between  $K(\tau,N)$ and $K_c(\tau,N)$ arises precisely from the sharp spectral edge. On the other hand, energy levels selected from an ensemble of levels that do not have a sharp edge such as those considered in \cref{sec:Numerical,sec:Origin} have
	\begin{align}
		Z(\tau,N) &=  \mathcal{O}\left(\frac{1}{N}\right), \\
		K(\tau,N) &\sim K_c(\tau,N) \sim N + \frac{2}{(\mu \tau)^2} + \ldots \label{eq:SFFPowerlaw}
	\end{align}
	i.e. both $K(\tau,N)$ and $K_c(\tau,N)$ have the same coefficient for the power-law scaling form. In summary, the sharp features in the DOS have no effect on the SFF but double the coefficient of the power-law scaling form in the connected SFF. Eliminating sharp features in the DOS has the desirable effect of leaving the power-law scaling form identical for the SFF and CSFF.
	
	\section{Sensitivity of $\chi(0,N)$ data on binning size} 
	\label{app:binning}
	The plots on the lower column of \cref{fig:r_chi0} were presented for a specific size of histogram bins used to produce $\chi(x,N)$ from which $\chi(0,N)$ was extracted. In \cref{fig:chi0_bins_critical}, we see that the curves are robust to a range of bin sizes (60/1500, 60/3000, 60/6000). In particular, the location of disorder strength where the curves for different system sizes cross, which estimates the transition between MBL and ergodic phases is unchanged. However, in order to use $N\chi(0,N)$ as an order parameter as we have done in \cref{fig:r_chi0}, we need to make sure that the size is not smaller than $\delta_L$ as defined in \cref{sec:level_repulsion} where it probes the finite-size level repulsions. 
	
	\section{Numerical parameters used to produce main text figures}
	\label{app:disorder}
	We list the details of various numerical parameters used to produce \cref{fig:SFF_numerics,fig:DOG,fig:r_chi0,fig:Poisson_XXZ_DOS_SFF,fig:chi_vs_bin,fig:QP_oscillations}. 
		\begin{table}[!htbp]
		\begin{center}
			\begin{tabular}{|c|c|c|}
				\hline
				20,000 & 30,000 & 30,000 \\
				\hline
				17,500 & 44,000 & 20,000.\\
				\hline
			\end{tabular}
		\end{center}
		\caption{Number of disorder samples used to produce plots in \cref{fig:SFF_numerics}.  \label{tab:SFF}}
	\end{table}

	\begin{table}[!htbp]
		\begin{center}
			\begin{tabular}{|c|c|c|}
				\hline
				30,000 & 30,000 & 30,000 \\
				\hline
				30,000 & 30,000 & 30,000.\\
				\hline
			\end{tabular}
		\end{center}
		\caption{Number of disorder samples used to produce plots in \cref{fig:DOG}. To produce the histograms, $1500$ bins are used for values of $x \in \left[-30,30\right]$  (also see \ref{app:binning} for details on sensitivity to binning). \label{tab:DOG}}
	\end{table}
	
\begin{table}[!htbp]
\begin{center}
	\begin{tabular}{|l|l|l|}
		\hline
		\begin{tabular}[c]{@{}l@{}}L=10: 50,000\\ L=12: 30,000\\ L=14: 20,000\\ L=16: 5,000\end{tabular} & \begin{tabular}[c]{@{}l@{}}L=8: 40,000\\ L=10: 30,000\\ L=12: 2,000\end{tabular} & \begin{tabular}[c]{@{}l@{}}L=10: 50,000\\ L=12: 30,000\\ L=14: 20,000\\ L=16: 5,000\end{tabular} \\ \hline
		\begin{tabular}[c]{@{}l@{}}L=10: 50,000\\ L=12: 30,000\\ L=14: 20,000\\ L=16: 5,000\end{tabular} & \begin{tabular}[c]{@{}l@{}}L=8: 40,000\\ L=10: 30,000\\ L=12: 2,000\end{tabular} & \begin{tabular}[c]{@{}l@{}}L=10: 50,000\\ L=12: 30,000\\ L=14: 20,000\\ L=16: 5,000\end{tabular} \\ \hline
	\end{tabular}
\end{center}
	\caption{Number of disorder samples used to produce plots in \cref{fig:r_chi0}. To produce the histograms, $1500$ bins are used for values of $x$ between $-30$ to $30$ (also see \ref{app:binning} for details on sensitivity to binning). \label{tab:r_chi0}}
\end{table}	

\begin{table}[!htbp]
\begin{center}
	\begin{tabular}{|l|}
		\hline
		\begin{tabular}[c]{@{}l@{}}L=10: 20,000\\        L=12: 20,000\\        L=14: 20,000 \\
			L=16: 10,000\end{tabular}                     \\ \hline
		\begin{tabular}[c]{@{}l@{}}       L=12: 20,000\\        L=14: 20,000\\        L=16: 10,000\end{tabular} \\ \hline
	\end{tabular}
\end{center}
	\caption{The number of disorder samples used to produce plots in \cref{fig:QP_oscillations}.}
\end{table} 
	
	\begin{table}[!htbp]
	\begin{center}
		\begin{tabular}{|c|c|c|}
			\hline
			NA & NA & 2,500 \\
			\hline
			40,000 & 40,000 & 17,500.\\
			\hline
		\end{tabular}
	\end{center}
	\caption{Number of disorder samples used to produce plots in \cref{fig:Poisson_XXZ_DOS_SFF}. The entries marked NA correspond to exact analytical plots. \label{tab:Poisson_XXZ_DOS_SFF}}
\end{table}

\begin{table}[!htbp]
\begin{center}
	\begin{tabular}{|l|}
		\hline
		\begin{tabular}[c]{@{}l@{}}L=10: 50,000\\        L=12: 30,000\\        L=14: 20,000\end{tabular}                     \\ \hline
		\begin{tabular}[c]{@{}l@{}}L=10: 50,000\\        L=12: 30,000\\        L=14: 20,000\\        L=16: 5,000\end{tabular} \\ \hline
	\end{tabular}
 \end{center}
\caption{The number of disorder samples used to produce plots in \cref{fig:chi_vs_bin}.}
\end{table} ~

	\newpage
	\bibliography{references}{}
	
\end{document}